\documentclass[aps,pra,twocolumn,floats,superscriptaddress,tighten,letterpaper,eqsecnum]{revtex4-1}
\usepackage{amssymb}
\usepackage{amsbsy}
\usepackage{amsmath}
\usepackage{graphicx}
\usepackage{graphics}
\usepackage{setspace}
\usepackage{array}
\usepackage{color}
\usepackage{xcolor}
\usepackage{fontenc}
\usepackage{textcomp}
\usepackage{rotating}
\usepackage{bm}
\usepackage{subfigure}
\usepackage{hyperref}
\hypersetup{pdfpagemode=UseNone}
\usepackage{empheq}
\usepackage{chngcntr}

\DeclareMathOperator{\diag}{diag}

\newcommand{\kc}{\msf{k}}

\newcommand{\e}{\varepsilon}

\newcommand{\vex}[1]{\bm{\mathrm{#1}}}
\newcommand{\Nabla}{\bm{\nabla}}

\newcommand{\pup}[1]{{\scriptscriptstyle{({#1})}}}
\newcommand{\puprm}[1]{{\scriptscriptstyle{(\mathrm{{#1}})}}}
\newcommand{\pupsf}[1]{{\scriptscriptstyle{(\mathsf{{#1}})}}}

\newcommand{\msf}[1]{\mathsf{#1}}
\newcommand{\ket}[1]{| {#1} \rangle}
\newcommand{\bra}[1]{\langle {#1} |}
\newcommand{\braless}[1]{\left\langle {#1} \right.}
\newcommand{\tauh}{\hat{\tau}}
\newcommand{\kaph}{\hat{\kappa}}
\newcommand{\sigh}{\hat{\sigma}}
\newcommand{\sigb}{\hat{\bm{\sigma}}}

\newcommand{\T}{\mathsf{T}}
\newcommand{\Mp}{\hat{M}_{\mathsf{P}}}
\newcommand{\Mps}{\hat{M}^{\pupsf{S}}_{\mathsf{P}}}
\newcommand{\Mt}{\hat{M}_{\mathsf{T}}}
\newcommand{\Ms}{\hat{M}_{\mathsf{S}}}
\newcommand{\hs}{\hat{h}}

\newcommand{\parr}{\partial}
\newcommand{\parb}{\bar{\partial}}

\newcommand{\bsub}{\begin{subequations}}
\newcommand{\esub}{\end{subequations}}

\newcommand{\gh}{\hat{\gamma}}
\newcommand{\ebulk}{E_{\mathsf{bulk}}}
\newcommand{\nimp}{n_{\mathsf{imp}}}
\newcommand{\ktf}{k_{\mathsf{TF}}}

\newcommand{\hsS}{\hat{h}_{\scriptscriptstyle{\mathsf{S}}}}

\newcommand{\Mts}{\hat{M}^{\pupsf{S}}_{\mathsf{T}}}
\newcommand{\Mss}{\hat{M}^{\pupsf{S}}_{\mathsf{S}}}

\newcommand{\te}{\tilde{\e}}

\newcommand{\intl}[1]{\int\limits_{#1}}
\newcommand{\nm}[1]{{\mathcal{N}_{{#1}}}}

\newcommand{\alp}{\hat{\alpha}}
\newcommand{\ti}{\hat{t}}
\newcommand{\ba}{\bar{a}}

\newcommand{\voo}{v_{11}}
\newcommand{\dvoo}{\delta v_{11}}
\newcommand{\vtt}{v_{22}}
\newcommand{\dvtt}{\delta v_{22}}
\newcommand{\vot}{v_{12}}
\newcommand{\vto}{v_{21}}

\begin{document}
\title{Criticality Across the Energy Spectrum from Random, Artificial Gravitational Lensing in Two-Dimensional Dirac Superconductors}
\author{Sayed Ali Akbar Ghorashi}
\thanks{These authors contributed equally.}
\affiliation{Department of Physics, William \& Mary, Williamsburg, Virginia 23187, USA}
\author{Jonas F. Karcher}
\thanks{These authors contributed equally.}
\affiliation{Institut f\"ur Nanotechnologie, Institut f\"ur QuantenMaterialien und Technologien and Institut f\"ur Theorie
der Kondensierten Materie, Karlsruhe Institute of Technology, 76021
Karlsruhe, Germany}
\author{Seth M. Davis}
\affiliation{Department of Physics and Astronomy, Rice University, Houston, Texas 77005, USA}
\author{Matthew S.\ Foster}
\affiliation{Department of Physics and Astronomy, Rice University, Houston, Texas 77005, USA}
\affiliation{Rice Center for Quantum Materials, Rice University, Houston, Texas 77005, USA}
\date{\today\\}

\newcommand{\be}{\begin{equation}}
\newcommand{\ee}{\end{equation}}
\newcommand{\bea}{\begin{eqnarray}}
\newcommand{\eea}{\end{eqnarray}}
\newcommand{\h}{\hspace{0.30 cm}}
\newcommand{\vs}{\vspace{0.30 cm}}
\newcommand{\n}{\nonumber}

\begin{abstract}
We numerically study weak, random, spatial velocity modulation [``quenched gravitational disorder'' (QGD)] 
in two-dimensional massless Dirac materials. 
QGD couples to the spatial components of the stress tensor; 
the gauge-invariant disorder strength is encoded in the quenched curvature.
Although expected to produce negligible effects, wave interference due to 
QGD transforms all but the lowest-energy states into a quantum-critical ``stack'' with universal, energy-independent spatial fluctuations. 
We study five variants of velocity disorder, incorporating three different local deformations of the Dirac cone:
flattening or steepening of the cone,
pseudospin rotations, 
and
nematic deformation (squishing) of the cone.
QGD should arise for nodal excitations in the $d$-wave cuprate superconductors (SCs), due 
to gap inhomogeneity. Our results may explain the division between low-energy ``coherent'' 
(plane-wave-like) and finite-energy ``incoherent'' (spatially inhomogeneous) excitations in the SC and 
pseudogap regimes.
The model variant that best matches the cuprate phenomenology possesses 
quenched random pseudospin rotations and nematic fluctuations. 
This model variant and another with pure nematic randomness exhibit a robust energy 
swath of stacked critical states, the width of which increases with increasing disorder strength. 
By contrast, quenched fluctuations that isotropically flatten or steepen the Dirac cone 
tend to produce strong disorder effects, with more rarified wave functions at low- and high-energies. 
Our models also describe the surface states of class DIII topological SCs.
\end{abstract}

\maketitle

\section{Introduction}

Understanding the interplay between strong correlations and quenched disorder in low-dimensional 
superconductors remains one of the key challenges in condensed matter physics \cite{TrivediReview}.
While pair-breaking due to elastic impurity scattering is detrimental to superconductivity, spatial inhomogeneity can 
locally \emph{enhance} pairing. 
Near a superconductor-insulator transition, the latter can occur via the spatial accumulation of Cooper pairs, 
due to the \emph{multifractal} rarification of single-particle eigenstates \cite{Feigelman07,Feigelman10,Foster12,Foster14,Burmistrov15}.
Multifractality arises in quantum-critical states due to wave interference induced by multiple impurity scattering.
Two-dimensional (2D) multifractal superconductivity has very recently come to the fore in studies of transition metal dichalcoginides 
\cite{Ji19,Ugeda18,Yeom19}. 

An enduring mystery is the role of spatial inhomogeneity in the high-$T_c$ cuprate superconductors. 
These materials are characterized by two different energy scales in the underdoped regime:
the pseudogap energy $\Delta_1$ and the smaller pairing energy $\Delta_0$; the former (latter) increases
(decreases) with increased underdoping \cite{Sawatzky08}.   
Maps of $\Delta_1(\vex{r})$ obtained via STM demonstrate strong nanoscale spatial inhomogeneity 
that increases with underdoping \cite{Davis01,Davis02,Davis05-a,Davis05-b,Davis08,DavisReview}.
A key observation in STM scans of the local density of states (LDOS) is that the lower
energy scale $\Delta_0$ appears to divide fermionic excitations into two distinct classes.
States with energies smaller than $\Delta_0$ behave like dispersive, ``coherent'' Bogoliubov-de Gennes 
quasiparticles, showing robust quasiparticle interference. 
States with energies above $\Delta_0$ are instead termed ``incoherent,'' as they exhibit strong spatial 
fluctuations that vary little with energy \cite{Davis08,DavisReview}. 
Renewed urgency for understanding spatial inhomogeneity comes from Ref.~\cite{Welp18},
which reported an increase in $T_c$ with increasing disorder. 

Much recent work on the strange physics of the cuprates 
invokes gravitational descriptions of quantum criticality \cite{Sachdev19}.
In this paper, we uncover a very different role possibly played by quantum criticality in 
these materials, due to a different type of ``gravitational'' physics. 
We show that the dichotomy between plane-wave-like, low-energy quasiparticle states and
strongly inhomogeneous, finite-energy states can be reconciled in a simple
model of noninteracting nodal Dirac quasiparticles, subject to random velocity disorder. 
Formally this can be cast as the effect of static spacetime curvature \cite{Ramond,Volovik,Juan12}
[``quenched gravitational disorder'' (QGD)]. 
For massless Dirac fermions, propagation is analogous to the lensing of 
starlight through a fixed, randomly gravitating spacetime \cite{Carroll}. 
The gravitational language precisely defines the gauge-invariant content of the 
disorder, through the induced curvature.

\begin{figure}[b!]
\centering
\includegraphics[width=0.45\textwidth]{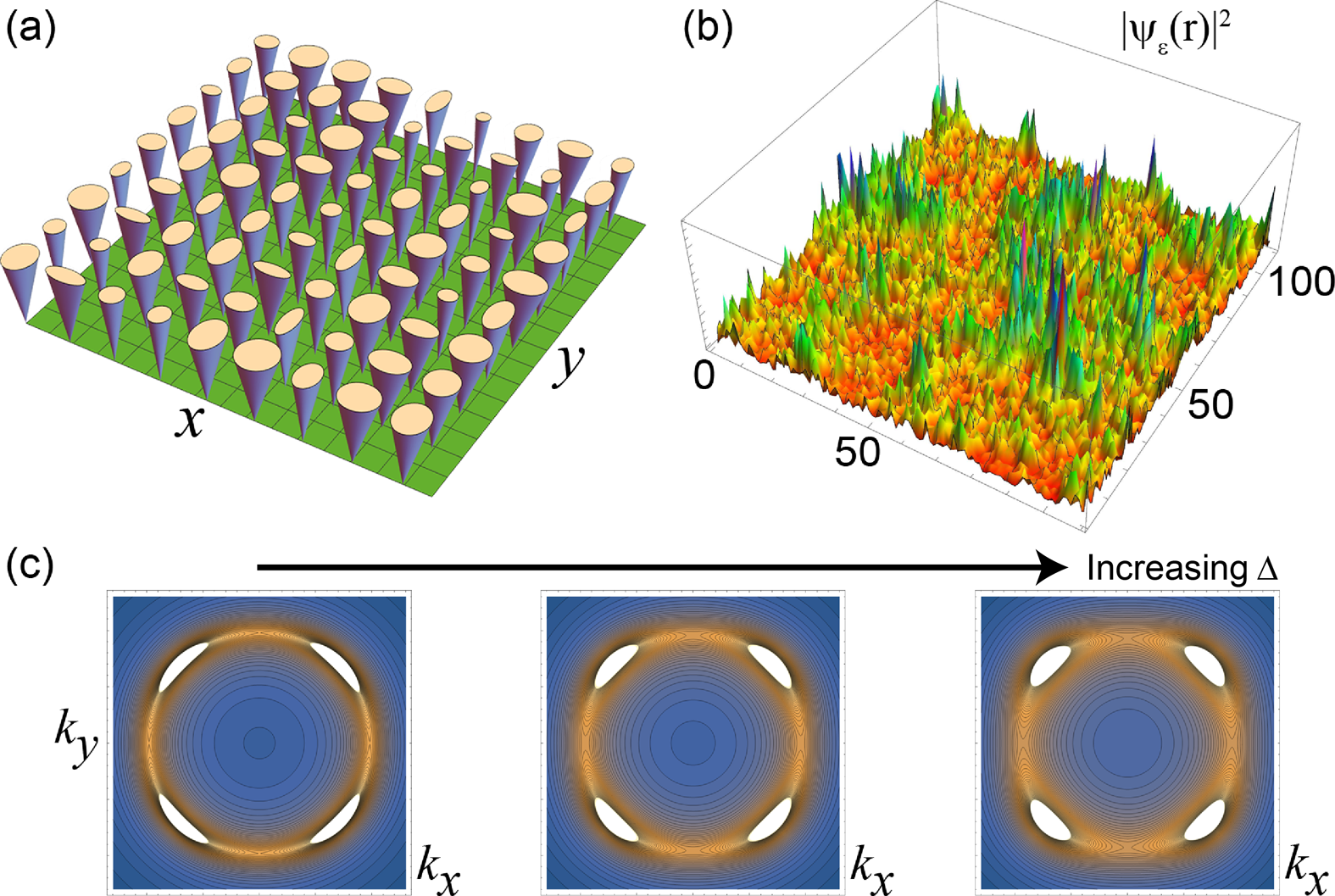}
\caption{
(a) Cartoon depicting random spatial modulation of the velocity [``quenched gravitational disorder'' (QGD)]
for 2D massless Dirac carriers. 
(b) Plot of the probability density $|\psi_\e(\vex{r})|^2$ for a \emph{critical state} wave function, 
representative of the ``stack'' of states found throughout the energy spectrum with QGD---see 
Figs.~\ref{Fig--Maps(b)}, \ref{Fig--Maps(e)}, \ref{Fig--DOS(b)}, and \ref{Fig--DOS(e)}.
(c) The nodal excitations of the $d$-wave cuprates in the superconducting and pseudogap phases should realize Dirac 
carriers with QGD. 
Plotted is the inverse dispersion $1/E(\vex{k})$ for different strengths of the order parameter amplitude $\Delta$, 
where
$E(\vex{k}) \equiv \sqrt{\te^2(\vex{k}) + \Delta^2 \cos^2(2 \phi_k)}$
describes quasiparticle excitations in a 2D $d$-wave superconductor 
\cite{ColemanBook}.
Here $\te(\vex{k})$ is the bare dispersion measured relative to the Fermi energy,
and $\phi_k$ is the polar momentum angle.
The nodal Dirac cones are indicated here in white. 
Spatial inhomogeneity in the amplitude $\Delta(\vex{r})$ \cite{Davis05-a,Davis05-b}
modulates the Dirac cones along the direction tangent to the Fermi surface. 
}
\label{Fig--RVC}
\end{figure}

QGD is realized whenever Dirac carriers arise from a correlation gap that is spatially inhomogeneous. 
In a $d$-wave superconductor, spatial gap fluctuations modulate the nodal quasiparticle velocity, 
see Fig.~\ref{Fig--RVC}. In the context of conventional 2D Dirac materials such as graphene or the surface
states of 3D topological insulators, QGD has so far received little attention \cite{Juan12,Nakai14,DiazFernandez17,Jafari19}. 
This is because disorder can usually couple in a more relevant fashion, through gauge and mass potentials
\cite{Juan12,AltlandSimonsZirnbauer02,AleinerEfetov06,Evers2008}. These perturbations typically 
induce metallic or Anderson insulating behavior on the largest scales
\cite{Bardarson07,Nomura07,Nomura08,Evers2008}.

\begin{figure}
\centering
\includegraphics[width=0.49\textwidth]{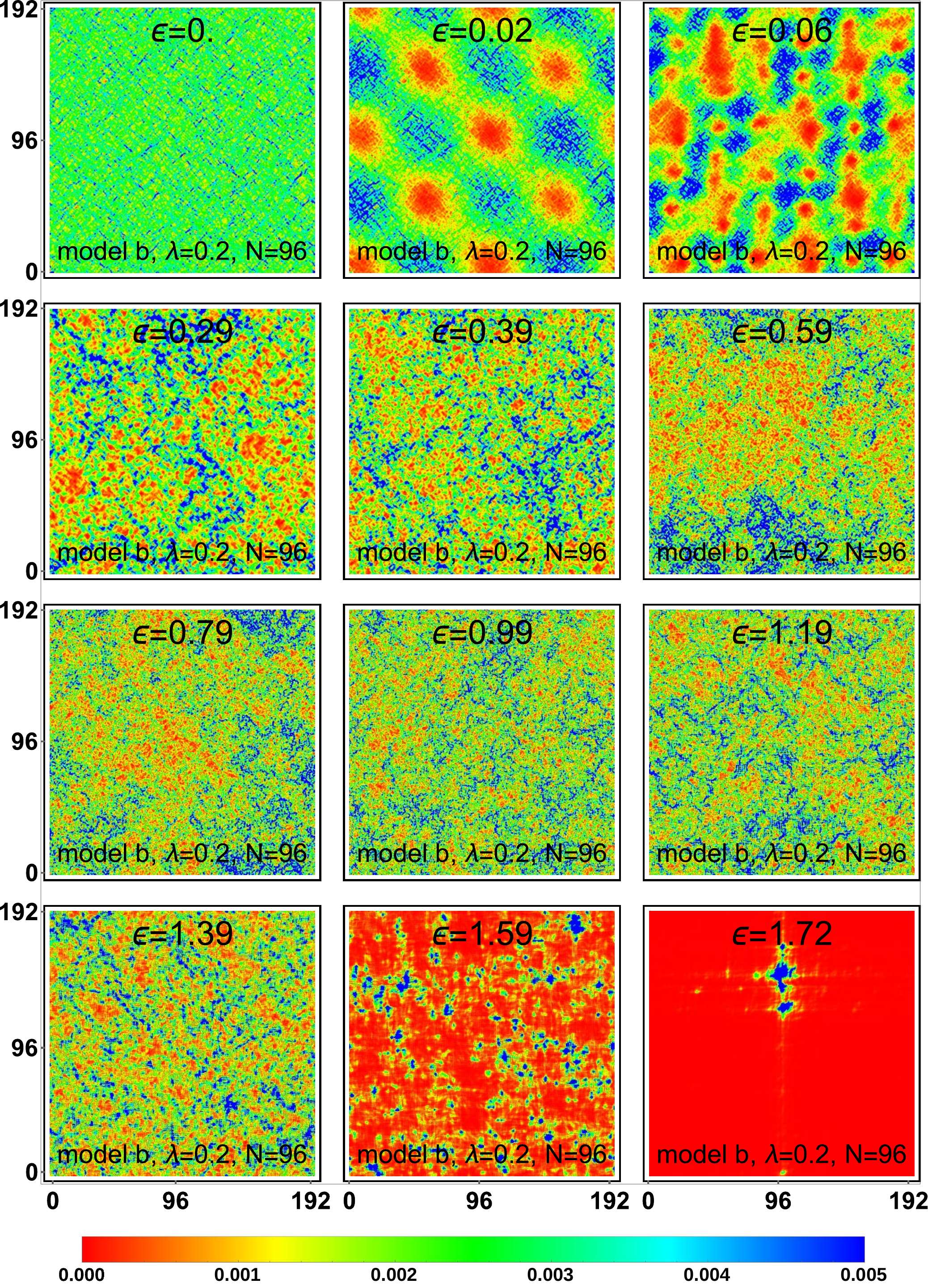}
\caption{
2D Dirac fermions with QGD: a spectrum dominated by critical states. 
The panels show 
position-space maps of $|\psi_\e(\vex{r})|^2$ for eigenstates $\{\psi_\e\}$ with energies $\e$ as quoted.
Energies (lengths) 
are measured in units of the momentum cutoff $\Lambda$
(inverse cutoff $2 \pi / \Lambda$), 
with the bare Dirac carrier velocity $v_0 = 1$.
Results here are presented for model (b) described below Eq.~(\ref{DiracRVH});
see Fig.~\ref{Fig--DOS(b)} for the density of states.
Although the lowest energy ($\e < 0.2$) states are plane-wave-like, most of the spectrum 
$(0.2 < \e < 1.5)$
consists of critical states. These are scale-invariant and extended (not Anderson localized), 
yet highly rarified \cite{Huckestein1995,Evers2008}. 
Critical states typically occur only with fine-tuning to a mobility edge, or to 
the quantum Hall plateau transition (QHPT) \cite{Evers2008,Huckestein1995}.
Recent work has demonstrated that ``stacks'' of critical QHPT states can 
nevertheless form at the 2D surface of 3D topological superconductors \cite{Chou2014,Ghorashi18,Sbierski}.
Results are obtained from exact diagonalization in momentum space over a $(2 N + 1)\times(2 N + 1)$ grid of 
momenta; here $N = 96$. 
Critical states are identified by their multifractal spectrum, see 
Fig.~\ref{Fig--Dirt_Series_Dq(b)}.  
The dimensionless disorder strength is $\lambda = 0.2$.
The criterion for ``strong disorder'' is $\lambda \gtrsim 0.393$ (Sec.~\ref{Sec:NumResults}).
}
\label{Fig--Maps(b)}
\end{figure}

\begin{figure}
\centering
\includegraphics[width=0.49\textwidth]{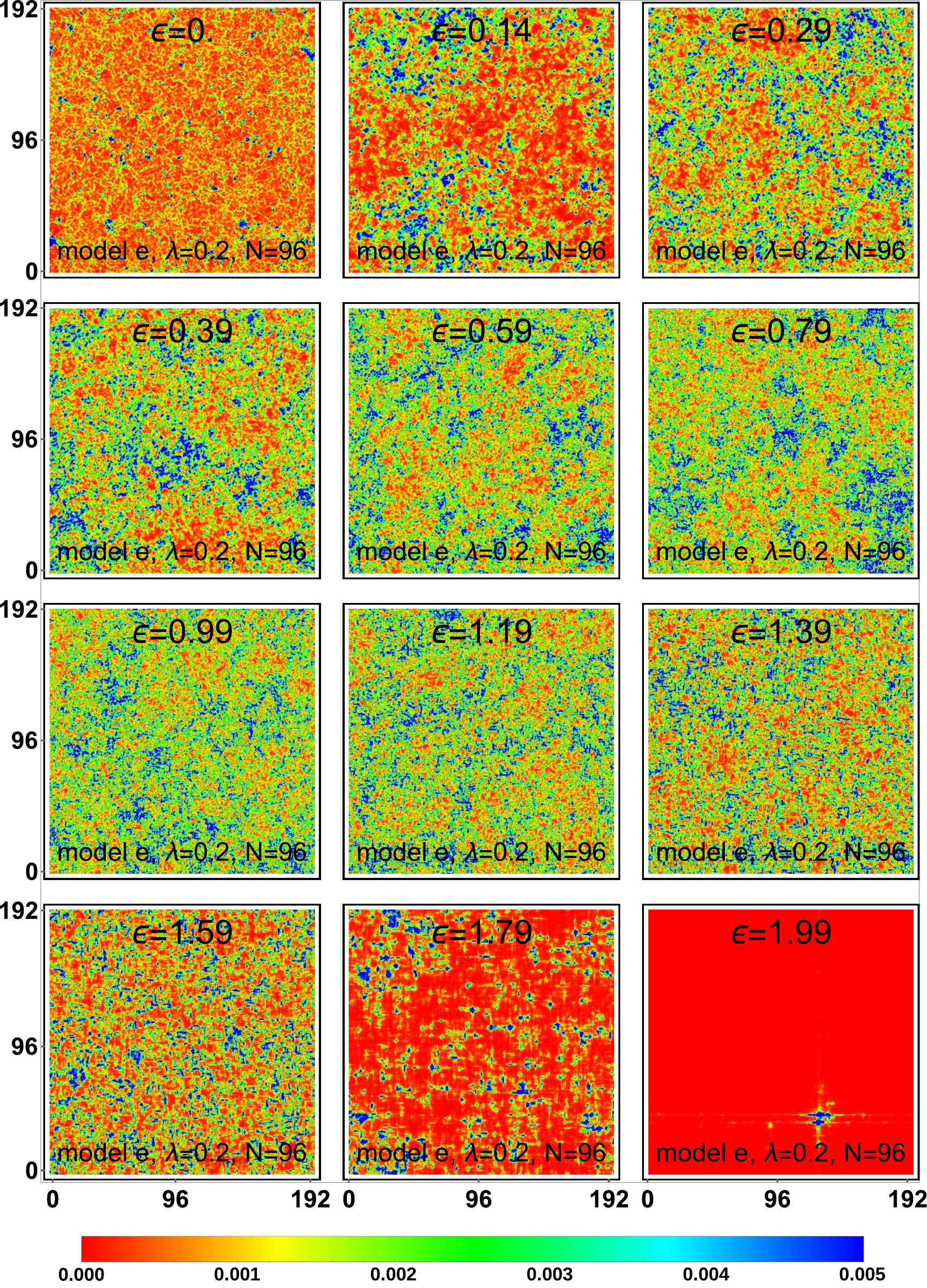}
\caption{
The same as Fig.~\ref{Fig--Maps(b)}, but for the generic model (e). 
See Fig.~\ref{Fig--DOS(e)} for the corresponding density of states 
and
Fig.~\ref{Fig--Dirt_Series_Dq(e)} for multifractal spectra. 
States with energies in the range $0.5 < \e < 1.5$ show critical stacking.
Unlike model (b), however, low-energy states are more strongly affected
by the disorder, showing supercritical (more rarified) fluctuations. 
In both Fig.~\ref{Fig--Maps(b)} and in this figure, Anderson localized states appear
deep in the Lifshitz tail, at high energies above the cutoff. 
}
\label{Fig--Maps(e)}
\end{figure}

In this paper, we use exact diagonalization to probe the effect of QGD on 2D Dirac
carriers. While the low-energy states near the Dirac point 
may be
only weakly affected,  
we find that \emph{most} of the energy spectrum converges into a ``stack'' of critical wave states.
These quantum-critical wave functions apparently exhibit universal multifractal LDOS fluctuations. 
A representative state is shown in Fig.~\ref{Fig--RVC}(b), while LDOS maps of states with different
energies appear in 
Figs.~\ref{Fig--Maps(b)} and \ref{Fig--Maps(e)}.

The Dirac model that we study is defined by the Hamiltonian
\begin{align}\label{DiracRVH}
	H
	=
	-
	\frac{1}{2}
	\sum_{a,b = 1,2}
	\int
	d^2\vex{r}
	\,
	v_{a b}(\vex{r})
	\left(
	\bar{\psi}
	i
	\sigh^a
	\!
	\stackrel{\leftrightarrow}{\parr_b}
	\!
	\psi
	\right),
\end{align}
where 
$\vex{r} = \{x_1,x_2\}$, 
$\psi = \psi_\sigma$ is a two-component spinor ($\sigma \in \{\uparrow,\downarrow\}$), 
$\{\sigh^{1,2}\}$ are Pauli matrices in the usual basis, 
$A \!\! \stackrel{\leftrightarrow}{\vspace{-12pt}\parr} \!\! B = A \parr B - (\parr A) B$,
and the QGD is encoded in the \emph{four} velocity components	
$\big\{	\voo(\vex{r}) \equiv 1 + \dvoo(\vex{r}),$ 
$	\vtt(\vex{r}) \equiv 1 + \dvtt(\vex{r}),$ 
$	\vot(\vex{r}),$		
$	\vto(\vex{r})	\big\}$.
The four potentials couple to the spatial components of the energy-momentum tensor $T^{a b}$. 
We consider five different variants of the model, with different combinations
of the random velocity perturbations. The variants are defined as follows,
\begin{enumerate}
\item[(a)]{Independent $\{\dvoo,\dvtt\}$, $\vot=\vto = 0$.
Local isotropic flattening or steepening of the Dirac cone and nematic (squishing) of the cone. 
}
\item[(b)]{Independent $\{\vot,\vto\}$, $\dvoo=\dvtt = 0$.
Local pseudospin rotations and nematic (squishing) of the Dirac cone. 
}
\item[(c)]{Independent $\{\dvoo = \dvtt,\;\vot = - \vto\}$.
Local isotropic flattening or steepening of the Dirac cone and pseudospin rotations. 
}
\item[(d)]{Independent $\{\dvoo = -\dvtt,\;\vot = \vto\}$.
Local nematic (squishing) of the Dirac cone.
}
\item[(e)]{Independent $\{\dvoo,\dvtt,\vot,\vto\}$.
The generic model.
}
\end{enumerate}
All nonzero random velocity components are taken to be short-range correlated,
characterized by a dimensionless disorder strength $\lambda$. 
The physical disorder variance is expressed as $\lambda$ times the square
of the inverse momentum cutoff (the disorder is formally irrelevant at the Dirac point---see below).

We find that critical-state stacking occurs for sufficiently weak disorder in 
all models (a)--(e).
Stacking \emph{improves} in models (b) and (d) with increasing disorder, 
i.e.\ more states become quantum-critical with identical, universal spectra 
throughout more of the spectrum. 
This is interesting because model (d) includes only \emph{pure nematic} deformations
of the Dirac cone. Nematicity has emerged as a key possible ingredient 
in the pseudogap phase of the cuprate superconductors \cite{NematicReview,PDWReview},
including quenched nematic fluctuations \cite{Davis2019}.
By contrast, models (a) and (c) (which incorporate local flattening and steepening of the cone)
are most susceptible to strong disorder effects, showing supercritical (more rarified)
states at low- and high-energies for increasing randomness. 
The generic model (e) shows intermediate behavior, including a robust energy swath with
critical stacking, but also supercritical behavior at low energies for stronger disorder. 
Critical stacking is exhibited for models (b) and (e) via the LDOS plots shown 
in Figs.~\ref{Fig--Maps(b)} and \ref{Fig--Maps(e)}.

Our results are very surprising, in that QGD is a \emph{strongly irrelevant} perturbation at low energies. 
QGD was instead expected to preserve the quasi-ballistic nature of the clean system, even away from the Dirac point \cite{Nakai14}. 
Moreover, critical wave states typically arise only at isolated energies, such as a mobility edge or at the plateau transition 
of a quantum Hall effect \cite{Huckestein1995,EversSQHE03,MirlinSQHE03,Evers2008}.
The population of critical states induced by QGD is compared to the total density of states in 
Figs.~\ref{Fig--DOS(a)}--\ref{Fig--DOS(e)} for models (a)--(e). 
Our calculations are performed in momentum space;
the largest size studied is a 193 $\times$ 193 grid. Since Dirac cones typically span 
$\sim 10 \,\%$ of the 2D Brillouin zone, this corresponds roughly to a (200 nm)$^2$ map. 
For models (b) and (d), more of the spectrum becomes more critical with increasing disorder 
(Figs.~\ref{Fig--DOS(b)} and \ref{Fig--DOS(d)}),
with very little dependence on the system size 
(Figs.~\ref{Fig--NSeriesDOCS(b)} and \ref{Fig--NSeriesDOCS(d)}).
The Dirac model with QGD in Eq.~(\ref{DiracRVH})
resides in class DIII; despite this, we find
no evidence for weak antilocalization at finite energy in any of the variants (a)--(e),
contrary to expectations based upon standard symmetry arguments \cite{Evers2008,Ghorashi18,Sbierski}.

Model (b) exhibits a phenomenology most similar to the LDOS maps observed in STM studies
of BSCCO \cite{Davis08,DavisReview}, with low-energy plane wave states and a robust 
quantum-critical stack at intermediate and higher energies, see Fig.~\ref{Fig--Maps(b)}. 
In addition to quenched nematic fluctuations, model (b) incorporates 
local rotations of the pseudospin $\{\sigh^1,\sigh^2\}$ relative to the 
coordinate axes $\{x_1,x_2\}$. 
As we review in Sec.~\ref{Sec:Review}, other types of disorder
afflicting $d$-wave quasiparticles
are predicted to induce different physics: 
(I) One class of (effectively topological \cite{Foster14}) perturbations should also 
produce critical scaling throughout the quasiparticle energy spectrum
\cite{Ostrovsky2007,Chou2014,Ghorashi18,Sbierski}.
In this scenario, though, zero-energy states are also predicted to be multifractal,
and the low-energy DOS $\nu(\e)$ would be enhanced via a sublinear power law, $\nu(\e \rightarrow 0) \sim |\e|^\delta$ with $\delta < 1$
\cite{Ludwig1994,Nersesyan1994,Mudry1996,Caux1996,Bhaseen2001,AltlandSimonsZirnbauer02}. 
(II) Disorder that induces generic internode scattering is instead predicted to Anderson localize the entire quasiparticle spectrum  
\cite{Senthil98,AltlandSimonsZirnbauer02}. 
Neither scenario (I,II) is consistent with STM data \cite{DavisReview}. 

QGD should also arise at the surface of a topological superconductor (TSC).
Due to ``topological protection,'' velocity modulation is the only allowed coupling
of charged impurities to the 2D massless Majorana fluid expected to form at the surface
of a class DIII TSC
\cite{TSCRev1,TSCRev2,TSCRev3,TSCRev4}
with winding number $|\nu| = 1$ \cite{Nakai14,Roy19}, as we show in Appendix \ref{Sec:3HeB}. 

The stack of critical finite-energy states found here is quite unusual, but not unprecedented. 
In Ref.~\cite{Ghorashi18}, we observed stacks of critical, class C spin quantum Hall plateau transition (QHPT) states
at finite energy in a surface model for a class CI TSC.
Numerical studies \cite{Chou2014,Sbierski} have found that the finite-energy surface states of class AIII TSCs 
sit at the class A integer QHPT \cite{Ludwig1994,Ostrovsky2007}. 	
We expect that the finite-energy critical states identified here 
correspond to a version of the class D \emph{thermal} QHPT \cite{D-1,D-2,D-3,D-4,D-5,D-6,D-7}.

\subsection{Outline}

This paper is organized as follows. 
In Sec.~\ref{Sec:Review}, we review results on impurity scattering in $d$-wave superconductors.
We derive the low-energy 4-node Dirac theory from a microscopic model.
We review how restrictions to
2- and 1-node models are \emph{effectively topological}, i.e.\ correspond to models for surface
states of bulk TSCs. 
We summarize results for
the thermal conductivity, density of states, and multifractal spectra of low- and finite-energy
wave functions. We discuss the quantum critical ``stacking'' of multifractal finite-energy states
and the connection to plateau transitions in quantum Hall effects 
that has recently been found to occur in these models \cite{Ghorashi18,Sbierski}.  

In Sec.~\ref{Sec:Results} we present the main numerical results of this paper
for the five different variants (a)--(e) of the model defined by Eq.~(\ref{DiracRVH}).
Our results consist of the density of states and multifractal exponents throughout the energy spectrum, 
for a range of disorder strengths and system sizes. We contrast our results to conventional expectations, 
and discuss them in light of the multifractal stacking phenomenon. 

The appendices detail more technical considerations. 
In Appendix \ref{Sec:Gravity}, we show how velocity randomness is embedded in a generally covariant framework
for 2+1-D Dirac fermions propagating through curved spacetime.  
In Appendix \ref{Sec:3HeB} we show how electric potentials couple gravitationally 
to the single Majorana surface cone predicted to occur at the boundary of a class DIII TSC. 
Finally, in Appendix \ref{Sec:Sym} we present a symmetry analysis of the Dirac fermion with QGD,
and the possible connection to the thermal QHPT in class D.


\section{Disorder in 2D $d$-wave superconductors, and topological superconductor surface fluids \label{Sec:Review}}

In this section we review a microscopic model for elastic impurity scattering in 2D $d$-wave
superconductors. The generic model with elastic scattering between all four low-energy Dirac quasiparticle
nodes Anderson localizes at all energies. 
We then show how restrictions on internode scattering produce 
independent, \emph{topologically protected} sectors. The latter are equivalent to the 
surface states of TSCs in classes CI, AIII, and DIII. 
Finally, we review exact results
for the thermal conductivity, density of states, and multifractal spectrum of local
density of states fluctuations. 
With the exception of the recent spectrum-wide criticality numerical results obtained in Refs.~\cite{Ghorashi18,Sbierski} and in this paper, 
all of the following is well known \cite{AltlandSimonsZirnbauer02,SRFL08,Foster14}.

The discussion below is based on the linearized quasiparticle band structure in the vicinity 
of the Dirac nodes. For a recent numerical study of disorder effects in a lattice model for the cuprates, 
see e.g.\ Ref.~\cite{Zaanen18}.

Readers primarily interested in new results for the Dirac model with QGD can
skip this section, and proceed directly to Sec.~\ref{Sec:Results}.

\subsection{$d$-wave model}

A 2D spin-singlet superconductor has the Bogoliubov-de Gennes Hamiltonian
\begin{align}\label{BdG2D}
\begin{gathered}[b]
	\!\!
	\!\!\!\!
	H = 
	\intl{\vex{k}} \Psi^\dagger(\vex{k}) \, \hs(\vex{k}) \, \Psi(\vex{k}),
	\;
	\Psi^\T(\vex{k}) 
	\equiv
	\begin{bmatrix}
	c_\uparrow(\vex{k}) \, c_{\downarrow}^\dagger(-\vex{k})
	\end{bmatrix},
	\!\!\!\!
\\
	\hs(\vex{k})
	=
	\begin{bmatrix}
	\te(\vex{k}) 		& \Delta(\vex{k}) \\
	\Delta^*(\vex{k}) 	& - \te(-\vex{k})
	\end{bmatrix}.
\end{gathered}
\end{align}
Here $\Psi(\vex{k})$ is the Nambu spinor that annihilates spin-1/2 quasiparticles,
$\T$ denotes the matrix transpose,
$\te(\vex{k}) = \e(\vex{k}) - \mu$ is the bare energy dispersion relative to the chemical potential $\mu$,
$\Delta(\vex{k})$ is the mean-field BCS order parameter, 
and 
$\intl{\vex{k}} \equiv \int d^2\vex{k}/(2 \pi)^2$.

The two key symmetries we want to enforce are $T^2 = -1$ time-reversal and 
spin SU(2) symmetry, the combination of which places the system in the Altland-Zirnbauer class CI \cite{AltlandSimonsZirnbauer02,Evers2008}.
U(1) symmetry under $\hat{S}^z$-axis spin rotations is manifest in Eq.~(\ref{BdG2D}). 
A rotation around $\hat{S}^{x}$ or $\hat{S}^y$ generates a continuous particle-hole transformation in the Nambu language.
To ensure SU(2) invariance, it is sufficient to impose invariance under a $\pi$ $\hat{S}^x$-rotation,
which takes the form of a $P^2 = -1$ effective particle-hole symmetry. 
These are encoded via the transformations
\bsub
\begin{align}
T:&\,&
	\Psi(\vex{k}) \rightarrow&\, i \sigh^2 \, \left[\Psi^\dagger(\vex{k})\right]^\T, \quad i \rightarrow -i,
\\
P:&\,&
	\Psi(\vex{k}) \rightarrow&\, - i \sigh^2 \, \left[\Psi^\dagger(-\vex{k})\right]^\T, 
\end{align}
\esub
where the Pauli matrices $\{\sigh^{1,2,3}\}$ act on the particle-hole space grading the Nambu spinor. 
The $T$ ($P$) transformation is antiunitary (unitary) in second quantization. 
These translate into the following conditions on $\hat{h}(\vex{k})$, 
\bsub\label{BdG_Sym}
\begin{align}
T:&\,&
	- \sigh^2 \, \hs(\vex{k}) \, \sigh^2 =&\, \hs(\vex{k}), 
\\
P:&\,&
	- \sigh^2 \, \hs^\T(-\vex{k}) \, \sigh^2 =&\, \hs(\vex{k}).
\end{align}
\esub
Physical time-reversal invariance is thus represented by an \emph{effective} unitary chiral symmetry 
in first quantization. 
Imposing both $T$ and spin SU(2) implies that $\te(-\vex{k}) = \te(\vex{k})$, and that
$\Delta(\vex{k}) = \Delta^*(\vex{k}) = \Delta(-\vex{k})$.

\begin{figure}[t]
\centering
\includegraphics[width=0.24\textwidth]{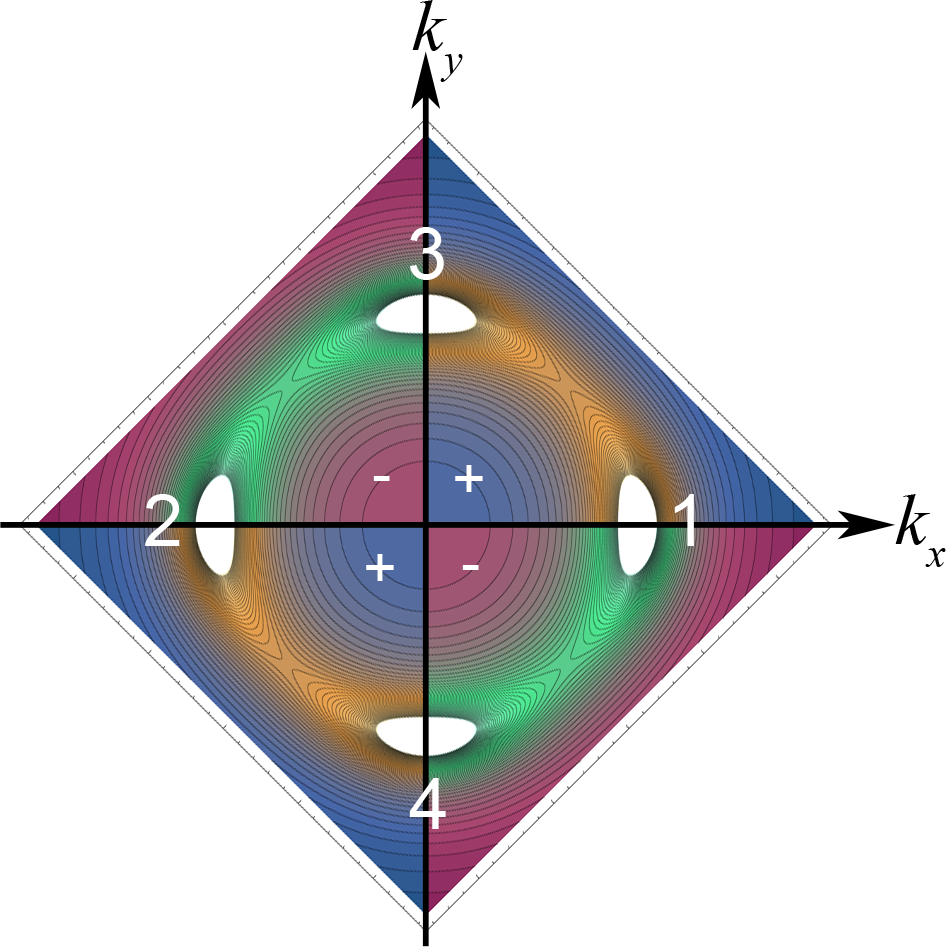}
\caption{The Dirac nodes of a 2D $d$-wave superconductor occur 
in two partnered pairs $(1,2)$ and $(3,4)$. 
The partners of a pair are related by time-reversal and spin rotation
symmetries. 
Generic impurity scattering between all four nodes is believed
to Anderson localize quasiparticle states at all energies \cite{Senthil98,AltlandSimonsZirnbauer02}.
On the contrary, if impurity scattering is restricted to occur only 
(i) between partners of a pair $(1 \Leftrightarrow 2)$ or $(3 \Leftrightarrow 4)$, 
or 
(ii) within each node separately,
then the model decouples (``fractionalizes'') into multiple \emph{topologically protected}
sectors \cite{SRFL08,Foster14}. These topologically protected sectors are equivalent to surface state theories
of bulk (3D) TSCs, in classes CI, AIII, or DIII (see text).  
}
\label{Fig--d-wave}
\end{figure}

For a $d_{x^2 - y^2}$ superconductor, we take $\Delta(\vex{k}) = \Delta_0 \, k_x k_y / k_F^2$,
where $\Delta_0$ is a real amplitude, $k_F$ is the Fermi wave vector, and $(k_x,k_y)$ measure
momentum along the nodal directions, see Fig.~\ref{Fig--d-wave}.  
Linearizing the Hamiltonian in the vicinity of each node gives an effective low-energy,
anisotropic Dirac theory,
\begin{align}\label{dwaveDirac}
	H
	=
	\intl{\vex{k}}
	\psi^\dagger(\vex{k})
	\,
	\hs_0(\vex{k})
	\,
	\psi(\vex{k}).
\end{align}
Here $\psi(\vex{k})$ is an 8-component spinor with long wavelength momentum $\vex{k}$.
The field $\psi$ is formed from the direct sum of $\Psi(\vex{k}_i + \vex{k})$ evaluated
in the vicinity of the four nodes $i \in \{1,2,3,4\}$ depicted in Fig.~\ref{Fig--d-wave},
located at $\{\vex{k}_i\}$. 
Let the Pauli matrices $\{\tauh^{1,2,3}\}$ act on the partners of a pair of nodes related by 
time-reversal, i.e.\ the matrix $\tauh^1$ exchanges states between nodes $(1 \Leftrightarrow 2)$ and $(3 \Leftrightarrow 4)$
in Fig.~\ref{Fig--d-wave}. Let the Pauli matrices $\{\kaph^{1,2,3}\}$ act on \emph{pairs} of nodes,
i.e.\ the matrix $\kaph^1$ exchanges $(1 \Leftrightarrow 3)$ and $(2 \Leftrightarrow 4)$.
The symmetries in Eq.~(\ref{BdG_Sym}) become 
\bsub\label{BdG_Sym_Dirac}
\begin{align}
T:&\,&
	- \sigh^2 \, \hs \, \sigh^2  =&\, \hs, 
\\
P:&\,&
	- \sigh^2 \, \tauh^1 \, \hs^\T \, \sigh^2 \, \tauh^1   =&\, \hs,
\end{align}
\esub
The Bogoliubov-de Gennes Hamiltonian takes the Dirac form
\begin{align}\label{h0Def}
	\hs_0
	=&\,
	\hat{P}_+
	\left[v_f \, \alp_1 (-i \parr_x) + v_\Delta \, \alp_2 (-i \parr_y)\right]
\nonumber\\
	&\,
	+
	\hat{P}_-
	\left[v_f \, \alp_1 (-i \parr_y) + v_\Delta \, \alp_2 (-i \parr_x)\right],
\end{align}
where $\hat{P}_{\pm} \equiv (1/2)(\hat{1}\pm\kaph^3)$ projects onto nodal pair ($1,2$) ($+$) or ($3,4$) ($-$),
and where 
the Clifford algebra matrices are 
$
	(\alp_1,\alp_2) = (\sigh^3\tauh^3,\sigh^1\tauh^3).
$
In Eq.~(\ref{h0Def}), $v_F$ is the bare Fermi velocity, while the perpendicular dispersion 
$v_\Delta \simeq \Delta_0/k_F$ arises from the pairing. 

Non-magnetic impurities can couple to the low-energy theory via local, Hermitian fermion bilinear operators.
The most relevant of these in the RG sense have no derivatives, and take the form
$
	\mathcal{O}^{\pup{\alpha\beta\gamma}}(\vex{r})
	\equiv
	\psi^\dagger(\vex{r})
	\,
	\sigh^\alpha
	\,
	\tauh^\beta
	\,
	\kaph^\gamma
	\,
	\psi(\vex{r}),
$
where $\alpha,\beta,\gamma \in \{0,1,2,3\}$ and $\sigh^0 = \tauh^0 = \kaph^0 = \hat{1}$ (the identity matrix). 
Imposing time-reversal and spin SU(2) symmetries from Eq.~(\ref{BdG_Sym_Dirac}) reduces the number
of allowed Hermitian bilinears to 20. This is consistent with the random matrix classification 
for class CI, where an $8\times8$ matrix is formed from the generators of Sp$(8)/$U$(4)$.
We can parameterize the 20 allowed perturbations via intrapair and interpair scattering Hamiltonians $\hat{h}_{1,2}$. 
Pure intrapair [$(1 \Leftrightarrow 2)$, $(3 \Leftrightarrow 4)$] scattering is encoded in
\begin{align}\label{h1Def}
	\hs_1
	=&\,
	\hat{P}_+
	\left[
	\alp_1
	\,
	A_{x,i}^\pup{+}(\vex{r}) 
	+
	\alp_2
	\,
	A_{y,i}^\pup{+}(\vex{r}) 
	\right]
	\ti^i
\nonumber\\
	&\,
	+
	\hat{P}_-
	\left[
	\alp_1
	\,
	A_{y,i}^\pup{-}(\vex{r}) 
	+
	\alp_2
	\,
	A_{x,i}^\pup{-}(\vex{r}) 
	\right]
	\ti^i,
\end{align}
where the repeated index $i$ is summed over $\{1,2,3\}$. 
The intrapair scattering terms take the form of SU(2) gauge potentials $\vex{A}_i^{\pup{\pm}}(\vex{r}) \, \ti^i$,
where the SU(2) generators are 
$
	\ti^i = \{\sigh^2 \tauh^1, \sigh^2 \tauh^2, \tauh^3\}. 
$
The remaining 8 allowed perturbations scattering between pairs of nodes,
and can be expressed through the Hamiltonian 
\begin{align}\label{h2Def}
	\hs_2
	=&\,
	\left[
	\kaph^1 
	\,
	B_{\ba,i}(\vex{r})\,\ti^i
	+
	\kaph^2
	\,
	C_{\ba}(\vex{r})
	\right]
	\alp_{\ba},
\end{align}
where repeated indices $i \in \{1,2,3\}$ and $\ba \in \{1,2\}$ are summed. 
In contrast to the intrapair scattering encoded in $\hs_1$, the interpair scattering mediated by $B_{\ba,i}(\vex{r})$ and $C_{\ba}(\vex{r})$ take
the form of scalar potentials and mass terms. 
The latter can be used to open up a gap without breaking $T$; this immediately implies that the generic $4$-node Hamiltonian
\begin{align}\label{4node}
	\hs
	\equiv
	\hs_0 + \hs_1 + \hs_2
\end{align}
cannot describe the isolated surface states of a strong TSC or insulator.  

It is possible to derive the strengths and correlations of all 20 disorder potentials $\{A_{\ba,i}^{\pup{\pm}},B_{\ba,i},C_{\ba}\}$
in the effective low-energy field theory from microscopic perturbations of the original model in Eq.~(\ref{BdG2D}) \cite{AltlandSimonsZirnbauer02},
but we will not pursue this here. In fact, our main interest will be not in these potential perturbations, but in the weaker ``quenched gravitational disorder''
(QGD, nodal velocity randomness), which we argue below should be present in the cuprate superconductors 
\cite{DavisReview}. 
In Appendix \ref{Sec:3HeB}, we provide a derivation of QGD from the coupling of the electric potential to Majorana surface states 
in a microscopic model of a class DIII TSC.

\subsection{Topological restrictions and TSC surface states: Localization, quantum criticality, and physical properties \label{Sec:DescModels}}

\subsubsection{Four coupled nodes: Spectrum-wide Anderson localization \label{Sec:CI-4}}

Quantum wave interference induced by generic elastic scattering between all four nodes in the model described by 
Eqs.~(\ref{h0Def})--(\ref{4node})
is expected to Anderson localize all quasiparticle states, at both zero and finite single-particle energies.

The localization near zero energy is understood as follows.
The Dirac model described by Eq.~(\ref{4node}) can be averaged over all 20 disorder potentials using the replica or supersymmetry (SUSY)
trick. For Gaussian white-noise distributions, the model flows to strong coupling under the perturbative renormalization group \cite{AltlandSimonsZirnbauer02}. 
One then expects that the system can be described by a replicated or SUSY nonlinear sigma model in class CI. 
Although this sigma model admits a Wess-Zumino-Novikov-Witten (WZNW) term, it can be shown that no such term arises in the full four-node theory 
(e.g., it is forbidden by average parity invariance \cite{AltlandSimonsZirnbauer02}; moreover, the WZNW model turns out to be topological \cite{SRFL08},
as reviewed below). 
Without the WZNW term, the 2D class CI sigma model also
flows to strong coupling, interpreted as the tendency towards Anderson localization \cite{Senthil98}. 

Anderson localization is anticipated at finite energy as well in the four-node model.
The standard argument (which, however, needs to be revised for topological models as we emphasize below) goes as follows.
At zero energy, 7 of the 10 Altland-Zirnbauer classes are characterized by a special particle-hole or chiral symmetry. 
This is true for all classes that describe spin-1/2, time-reversal-invariant superconductors in classes CI, AIII, and DIII \cite{SRFL08,Foster14}.  
This symmetry also relates states at positive and negative energies, but does not tell us anything about the character
of the states at some particular fixed energy $\e \neq 0$. On the other hand, for quasiparticle states in a superconductor, 
one can always consider $|\e| \gg \Delta_0$, where $\Delta_0$ is the pairing energy. 
Then the quasiparticle states should reside in the standard Wigner-Dyson class of the 
normal metal that hosts the superconductivity. Since $\e = 0$ is the only symmetry-distinguished energy,
one concludes that \emph{all} states with $\e \neq 0$ in the infinite-volume limit must reside in the 
orthogonal, unitary, or symplectic Wigner-Dyson class. 
For the four-node class CI model, the appropriate Wigner-Dyson class is the orthogonal metal class AI,
which also preserves $T$ and spin SU(2) symmetry. This class is believed to always localize in 2D \cite{Evers2008}. 

Localization of all quasiparticle states is not borne out by experiments in the 
high-$T_c$ cuprate superconductors, at or below optimal hole doping. 
First of all, localization is at variance with superconductivity itself,
since $T_c$ is not protected by Anderson's theorem for non-$s$-wave pairing \cite{AltlandSimonsZirnbauer02}. 
Second, STM data taken at energies less than a characteristic scale $\Delta_0$ (which is always below
the pseudogap energy $\Delta_1$ on the underdoped side) show robust quasiparticle interference, a sign that 
there is just enough internode scattering to reveal the clean dispersion, but not enough to induce localization
\cite{Davis08,DavisReview}. 

Finally, the experimentally measured low-temperature longitudinal thermal conductivity 
\cite{Taillefer00,Orenstein00}
is nonzero, and close to the universal (disorder-independent) theoretical result
\begin{align}\label{TC}
	\frac{\kappa}{T}
	=
	\frac{k_B^2}{3 \hbar}
	\left(\frac{v_F}{v_\Delta}+\frac{v_\Delta}{v_F}\right). 
\end{align}
This result was originally obtained via an approximate, self-consistent semiclassical calculation \cite{Lee93,Senthil98,LeeDurst00}.
Eq.~(\ref{TC}) is
better understood as the \emph{exact} result for the $T \rightarrow 0$ limit of the 
Landauer thermal conductivity in the \emph{clean}
four-node model, the analog (via Wiedemann-Franz) 
of the ``ballistic'' conductivity $\sigma^{xx} = 4 e^2/\pi h$ for pristine graphene
doped exactly to charge neutrality. The absence of impurity scattering combined with the vanishing
density of states produces a finite, universal result due to evanescent wave propagation \cite{Tworzydlo06}. 
A very special feature of the topological models reviewed below is that Eq.~(\ref{TC}) remains exact
in the presence of disorder \cite{Ludwig1994,Tsvelik95,Ostrovsky06,SRFL08}, and is even predicted to be 
independent of virtual interaction (Altshuler-Aronov) corrections in these special models \cite{Xie15}.
By contrast, the Anderson localized model in Eq.~(\ref{4node}) would have $\kappa/T \rightarrow 0$ 
as $T \rightarrow 0$.

\subsubsection{Two coupled nodes: class CI TSC surface states and spectrum-wide spin quantum Hall criticality \label{Sec:CI-2}}

If we restrict ourselves to intrapair scattering, such that the node pair (1,2) decouples from (3,4) (Fig.~\ref{Fig--d-wave}),
then $\hs = \hs_0 + \hs_1$ [Eqs.~(\ref{h0Def}) and (\ref{h1Def})]. Since the pairs are independent, we focus
on one pair (1,2). After a rescaling of $(x,y)$ and a basis change, the two-node Hamiltonian can be written as 
\begin{align}\label{CI-2}
	\hs_{{\mathrm{CI}_2}} = \sigb\cdot\left[-i \Nabla + \vex{A}_i(\vex{r}) \, \tauh^i\right],
\end{align}
where $i \in \{1,2,3\}$ and $\sigb \equiv \sigh^1 \, \hat{x} + \sigh^2 \, \hat{y}$.
Here $\hs_{{\mathrm{CI}_2}}$ is a $4 \times 4$ matrix differential operator 
acting in the composite (particle-hole)$\otimes$(valley) ($\sigma \otimes \tau$) space. 
Since nodes (1,2) are related by $T$, the model still resides in class CI. The symmetry
operations in Eq.~(\ref{BdG_Sym_Dirac}) become
\bsub\label{Sym_2node}
\begin{align}
T:&\,&
	- \sigh^3 \, \hs \, \sigh^3  =&\, \hs, \label{T2node}
\\
P:&\,&
	- \sigh^1 \, \tauh^2 \, \hs^\T \, \sigh^1 \, \tauh^2   =&\, \hs.
\end{align}
\esub

Given the (transformed) form of the Clifford algebra in Eq.~(\ref{CI-2}), 
the effective chiral/physical time-reversal symmetry condition in Eq.~(\ref{T2node}) is anomalous, 
and cannot be realized without fine-tuning in two spatial dimensions \cite{BernardLeClair02,SRFL08}. It is naturally realized
on the surface of a class CI TSC \cite{SRFL08,Schnyder09,Foster12,Foster14},
with minimal winding number $|\nu| = 2$. 
The sigma model takes the same form as the four-node model, except that it is now augmented with a WZNW term at level $k = 1$ 
\cite{Nersesyan1994,Mudry1996,Caux1996,Bhaseen2001,Foster14,Schnyder09}. 

The density of states $\nu(\e)$ exhibits critical scaling with energy $\e$ in the limit $|\e| \rightarrow 0$. 
In particular, 
\begin{align}\label{CI-DOS}
	\lim_{\e \rightarrow 0}
	\,
	\nu(\e) \sim |\e|^{x_1/z},
\end{align}
where $x_1/z = 1/7$ \cite{Nersesyan1994}. 
The zero-energy wave function $\psi_0(\vex{r})$ exhibits quantum critical fluctuations on all length scales; 
these are characterized by the multifractal spectrum $\tau(q)$ (for a review, see e.g.\ \cite{Evers2008}).
If the system has size $L \times L$, one divides this up into $N^2$ boxes of size $b$, with $N \equiv L/b$. 
Then one introduces the box probability $\mu_i \equiv \intl{b_i} d^2\vex{r} \, |\psi_0(\vex{r})|^2$,
where the integral is performed over the $i^{\mathrm{th}}$ box. 
The multifractal spectrum is defined via the scaling behavior of moments of the box probability,
\begin{align}\label{tau(q)Def}
	\sum_{i = 1}^{N^2} (\mu_i)^q \sim \left(\frac{b}{L}\right)^{\tau(q)}. 
\end{align}
Box probabilities can also be obtained by normalizing a spatial map of the local density of states
in an STM experiment. 
In the limit $L \rightarrow \infty$, $\tau(q)$ is self-averaging. 
For the topological class CI model in Eq.~(\ref{CI-2}), the spectrum is perfectly parabolic.
The exact result is  
\begin{align}\label{CI-MFC}
\begin{aligned}
	\tau(q) =&\, 2(q - 1) + \Delta(q),
\\
	\Delta(q) =&\, \theta \, q(1 - q), 
\quad 
	0 \leq |q| \leq q_c, \;\; q_c \equiv \sqrt{\frac{2}{\theta}}, 
\end{aligned}
\end{align}
with $\theta = 1/4$ \cite{Mudry1996,Caux1996}. 

Very recently, the authors considered the question of the \emph{finite-energy} states of 
the model in Eq.~(\ref{CI-2}). On one hand, the same argument presented in Sec.~\ref{Sec:CI-4} leads
to the conclusion that all finite-energy states should be Anderson localized in the orthogonal class AI;
this was the ``conventional wisdom'' \cite{Evers2008}. 
On the other hand, Eq.~(\ref{CI-2}) also describes a surface quasiparticle fluid that forms at the boundary 
of a bulk TSC \cite{SRFL08,Schnyder09},
protected by the anomalous form of time-reversal symmetry in Eq.~(\ref{T2node}). 
From this perspective, the
idea that only the zero-energy single-particle wave function $\psi_0(\vex{r})$ escapes 
Anderson localization appears very strange. Indeed, it would correspond to a very weak
form of ``topological protection,'' since in other topological phases such as quantum Hall liquids,
2D and 3D topological insulators, it is the entire \emph{band} of edge or surface excitations
that is protected from Anderson localization \cite{KaneHasan,TSCRev1}.

\begin{table*}[!ht]
\def\arraystretch{1.5}
\begin{tabular}{clclcl|llllllll}
\vspace{-6pt}
	\# of
&
\;
&
&
\;
&
	Effective
&
\;
&
\;
&
&
\;
&
&
\;
&
&
\;
&
\\
\vspace{-6pt}
	nodes
&
\;
&
	Class
&
\;
&
	winding
&
\;
&
\;
&
	$x_1/z$ 
&
\;
&
	$\theta(\e = 0)$
&
\;
&
	$\theta(\e \neq 0)$
&
\;
&
	Dirt type(s)
\\
	coupled
&
\;
&
&
\;
&
	\# $|\nu|$
&
\;
&
\;
&
&
\;
&
&
\;
&
&
\;
&
\\
\hline 
\hline
	4
&
\;
&
	CI
&
\;
&
	N/A
&
\;
&
\;
&
	1 \cite{SenthilFisher99} 
&
\;
&
	N/A
&
\;
&
	N/A
&
\;
&
	Vector, potential,
\\
&
\;
&
&
\;
&
&
\;
&
\;
&
&
\;
&
	(localized) \cite{Senthil98,AltlandSimonsZirnbauer02}
&
\;
&
	(localized) \cite{Evers2008}
&
\;
&
	mass \cite{AltlandSimonsZirnbauer02}
\\
\hline
	2
&
\;
&
	CI
&
\;
&
	2
&
\;
&
\;
&
	1/7 \cite{Nersesyan1994} 
&
\;
&
	1/4 \cite{Mudry1996,Caux1996}
&
\;
&
	$\simeq 1/8$ (SQHPT) \cite{Ghorashi18}
&
\;
&
	SU(2) vector
\\
\hline
	1
&
\;
&
	AIII
&
\;
&
	1
&
\;
&
\;
&
	${\frac{1 - \lambda_A}{1 + \lambda_A}}$ \cite{Ludwig1994} 
&
\;
&
	$\lambda_A$ \cite{Ludwig1994}
&
\;
&
	$\simeq 1/4$ (IQHPT) \cite{Chou2014,Sbierski}
&
\;
&
	U(1) vector
\\
\hline
	1
&
\;
&
	DIII
&
\;
&
	1
&
\;
&
\;
&
	1 (clean)
&
\;
&
	0 (clean) 
&
\;
&
	$\simeq 1/13$ (TQHPT?) Sec.~\ref{Sec:Results}
&
\;
&
	Velocity/QGD
\\
\hline \hline
\end{tabular}
\caption{ 
Key properties of the effective 2D dirty Dirac Bogoliubov-de Gennes Hamiltonian for quasiparticles
in a $d$-wave superconductor. The number of nodes coupled refers to elastic impurity scattering 
between and/or within nodes. The generic model is expected to Anderson localize at all energies.
The models restricted to disorder that couples only 2 or 1 node(s) are all effectively topological,
i.e.\ describe the surface states of 3D bulk TSCs. 
All of the restricted models have the low-energy thermal conductivity given by Eq.~(\ref{TC}),
independent of disorder \cite{Ludwig1994,Tsvelik95,Ostrovsky06} and interactions \cite{Xie15}. 
The exponent $x_1/z$ governs the low-energy scaling of the density of states [Eq.~(\ref{CI-DOS})].
The restricted models all exhibit ``multifractal stacking'' of quantum critical wave functions throughout
the energy spectrum. 
The parameter $\theta(\e)$ characterizes the multifractal spectrum of wave functions according
to Eq.~(\ref{CI-MFC}), near energy $\e$. 
The stacked critical states for classes CI and AIII apparently belong to the spin and
integer quantum Hall plateau transitions (SQHPT and IQHPT, respectively). 
In the last entry, TQHPT refers to the thermal quantum Hall plateau transition; 
see Appendix \ref{Sec:Sym} for a discussion. 
}~\label{TSCTable}
\end{table*}

The only alternative to localization 
was argued in Ref.~\cite{Ghorashi18} to be a ``stacking'' of identical, quantum critical wave functions
at all nonzero energies. It was argued that each such wave function should sit at the 
class C, \emph{spin} quantum Hall plateau transition (SQHPT) \cite{SQHP1,SQHP2,SQHP3}. 
The latter shares a few critical exponents with classical 2D percolation
\cite{SQHP2,EversSQHE03,MirlinSQHE03,Bhardwaj2015}, a logarithmic conformal field theory \cite{SQHP4,SQHP5}.  

The numerical results of Ref.~\cite{Ghorashi18} are consistent with the ``critical stacking''
scenario. Near zero energy, the DOS and multifractal spectrum confirm the predictions
of the WZNW theory [Eqs.~(\ref{CI-DOS}) and (\ref{CI-MFC}), with $x_1/z = 1/7$ and $\theta = 1/4$]. 
At intermediate energies, however, a wide swath of states is found to exhibit weaker universal 
multifractality, given approximately by Eq.~(\ref{CI-MFC}), but now with $\theta \simeq 1/8$.
The latter is consistent with the SQHPT \cite{EversSQHE03,MirlinSQHE03}. 
More states exhibit SQHPT phenomenology upon increasing the disorder strength or system size \cite{Ghorashi18}. 

Finally, we note that the results described above are clearly incompatible with
experiments in the cuprates. 
This is not surprising, because in the 2D $d$-wave model (Fig.~\ref{Fig--d-wave})
it is difficult to microscopically suppress scattering between node pairs
$(1,2) \Leftrightarrow (3,4)$, 
while simultaneously 
retaining significant internode scattering between partners ($1 \Leftrightarrow 2$) and ($3 \Leftrightarrow 4$). 
The low-energy density of states vanishes as a strongly sublinear power-law $\nu(\e) \sim |\e|^{1/7}$. 
The two-node model exhibits the \emph{strongest} multifractality at zero energy [Eq.~(\ref{CI-MFC}) with $\theta = 1/4$], and 
\emph{weaker} multifractality at finite energies [Eq.~(\ref{CI-MFC}) with $\theta \simeq 1/8$ \cite{Ghorashi18}].
These features are opposite the observations in STM on BSCCO, which show minimal spatial inhomogeneity in 
low-energy LDOS maps, with stronger inhomogeneity above the energy scale $\Delta_0$;
in addition, the low-energy DOS retains the linear character of the clean system 
\cite{DavisReview}.

\subsubsection{One node, vector potential disorder: class AIII TSC surface states and spectrum-wide quantum Hall criticality}

We can further restrict to pure intranode scattering.
The effective Hamiltonian is ``half'' of Eq.~(\ref{CI-2}), 
i.e.\ a single 2-component Dirac fermion subject to U(1) vector potential randomness,
\begin{align}\label{AIII-1}
	\hs_{\mathrm{AIII}_1} = \sigb\cdot\left[-i \Nabla + \vex{A}(\vex{r})\right].
\end{align}
Formally this single-node Hamiltonian is the same as the surface fluid of a class AIII
TSC with minimal winding number $|\nu| = 1$. 
The anomalous (topological) time-reversal symmetry is still encoded by Eq.~(\ref{T2node}).

The density of states scales as in Eq.~(\ref{CI-DOS}), 
with exponent \cite{Ludwig1994}
\begin{align}\label{AIII-DOS}
	{x_1}/{z} = (1 - \lambda_A)/(1 + \lambda_A).
\end{align}
Here $\lambda_A$ denotes the variance of the assumed white-noise vector disorder potential,
\[
	\overline{A_{\ba}(\vex{r}) \, A_{\ba'}(\vex{r'})} = \pi \, \lambda_A \, \delta_{\ba,\ba'} \, \delta^{\pup{2}}(\vex{r} - \vex{r'}). 
\]
At zero energy, the multifractal spectrum is given by Eq.~(\ref{CI-MFC}),
with \cite{Ludwig1994}
\begin{align}\label{AIII-MFC}
	\theta = \lambda_A. 
\end{align}
For a review of this model in the context of TSCs and its higher-winding number generalizations, see e.g.\ Ref.~\cite{Foster14}.  

At finite energy, the eigenstates of Eq.~(\ref{AIII-1}) form a stack of quantum-critical wave functions as in
the CI case. These were expected to reside at the ordinary class A integer quantum
Hall plateau transition (IQHPT) \cite{Ludwig1994,Ostrovsky2007}, and this result has been confirmed numerically in Refs.~\cite{Chou2014,Sbierski}. 
The IQHPT has an approximately parabolic multifractal spectrum as in Eq.~(\ref{CI-MFC}), with $\theta \simeq 1/4$ \cite{Huckestein1995,Evers2008}. 

The spectrum-wide ``stacked'' IQHPT multifractality is quite strong.
Unlike the $|\nu| = 2$ class CI model, the low-energy class AIII model predictions for the DOS and multifractal spectrum depend on 
the nonuniversal parameter $\lambda_A$. When the latter is strong enough to render the finite-energy states critical over a length
scale that is not too large, one would expect to see multifractality extending all the way down to zero energy [Eq.~(\ref{AIII-MFC})],
as well as a nonlinear enhancement of the DOS [Eqs.~(\ref{CI-DOS}) and (\ref{AIII-DOS})]. Neither of these features are seen 
in STM data on BSCCO \cite{DavisReview}.

\subsubsection{One node, gravitational disorder due to spatial gap inhomogeneity: class DIII TSC surface states and spectrum-wide criticality}

The simplest possible model neglects all forms of potential scattering. 
However, even in this case it is possible for disorder to produce a nonzero effect. 
In particular, a slow spatial modulation of $d$-wave the gap amplitude $\Delta_0 = \Delta_0(\vex{r})$
should induce spatial modulation of $v_\Delta \sim \Delta_0/k_F$ in Eq.~(\ref{h0Def}),
referred to in the sequel as QGD.  
Because the disorder couples to an operator with a spatial derivative, it is formally 
irrelevant in an RG sense at zero energy, in contrast with the potential perturbations in Eqs.~(\ref{h1Def}) and (\ref{h2Def}).
The latter are marginal (at tree level); these conclusions assume short-range correlated disorder. 
As a result, for sufficiently weak QGD in Eq.~(\ref{DiracRVH}),
the low-energy DOS $\nu(\e) \sim |\e|$ as in the clean limit, 
and the states near zero energy are not multifractal.
Although velocity disorder modifies the definition of the (spin) current operator, 
the low-temperature thermal conductivity should be unchanged from the clean Landauer result in Eq.~(\ref{TC}),
due to the irrelevance of disorder near $\e = 0$.  

The effects of such velocity disorder at finite $\e \neq 0$ 
are the main focus of this paper; the setup and results are discussed
in the next section. Formally, a single node with QGD resides in class DIII,
and could be realized as a dirty Majorana cone on the surface a bulk 
TSC (such as the candidate material Cu$_x$Bi$_2$Se$_3$ \cite{TSCRev1}). 
This is discussed in detail in Appendix \ref{Sec:3HeB}. 
A summary of the results discussed in this section appears in Table~\ref{TSCTable}.


\section{2D Dirac fermions with quenched gravitational disorder: Results \label{Sec:Results}}

\subsection{Model, formulations, and applications}

The single 2D Dirac fermion with quenched velocity disorder described by the Hamiltonian in Eq.~(\ref{DiracRVH}) 
can be associated to a (2+1)-D action
\begin{align}\label{DiracRV}
	\!\!\!\!
	S
	=
	\int
	d t
	\,
	d^2\vex{r}
	\!
	\left[
	\bar{\psi}
	\,
	i \parr_t
	\,
	\psi
	+
	\!\!
	\sum_{a,b = 1,2}
	\frac{v_{ab}(\vex{r})}{2}
	\!
	\left(
	\bar{\psi}
	i
	\sigh^a
	\!
	\stackrel{\leftrightarrow}{\parr_b}
	\!
	\psi
	\right)
	\right]\!\!.
	\!\!\!\!
\end{align}
The action can be interpreted in terms of fermions propagating through curved spacetime, with a static metric 
$g_{\mu \nu}(\vex{r})$ defined explicitly in terms of the velocity fields $\{v_{ab}\}$ in Appendix~\ref{Sec:Gravity}.
Eq.~(\ref{DiracRV}) obtains from $g_{\mu \nu}(\vex{r})$ and from the covariant action
\begin{align}
\label{DiracCurv}
	S 
	=&\,
	\int
	\sqrt{|g|}
	\,
	d^3 x
	\,
	\bar{\psi}
	\,
	E^\mu_{A}
	\,
	\gh^A
	\big(
	i
	\parr_\mu
	-
	{\textstyle{\frac{1}{2}}}
	\omega_\mu{}^{B C}
	\hat{S}_{B C}
	\big)
	\psi,
\end{align}
where $\mu \in \{t,x_1,x_2\}$ and $A,B,C \in \{0,1,2\}$; 
repeated indices are summed. 
Here $\sqrt{|g|} \, d^3 x$ is the volume measure,
$\{\gh^A\}$ are the gamma matrices, 
$E^\mu_{A}$ is the ``dreibein,'' 
$\omega_\mu{}^{B C}$ is the spin connection,
and 
$\hat{S}_{B C}$ generates local Lorentz
transformations \cite{Carroll}. 
Since the velocity modulation enters through the 
effects of spacetime curvature in Eq.~(\ref{DiracCurv}), 
we alternatively refer to this as 
``quenched gravitational disorder'' (QGD).
The gauge-invariant content of the disorder can be characterized
via the induced curvature, as shown in Appendix \ref{Sec:Gravity}.

The Dirac model in Eqs.~(\ref{DiracRVH}) and (\ref{DiracRV}) is noninteracting.
We can therefore alternatively cast the problem in terms of (2+0)-D action
designed to compute Green's functions at a given, fixed single particle energy $\e$. 
Using the chiral decomposition 
$\psi \equiv \begin{bmatrix} L & R \end{bmatrix}$,
$\bar{\psi} \equiv \begin{bmatrix} \bar{R} & \bar{L} \end{bmatrix}$,
we get a perturbed two-dimensional free-fermion conformal field theory
\begin{align}\label{SCFT}
	\!\!\!\!
	S
	=
	\int 
	d^2\vex{r}
	\left\{
		-\left[
			\eta^{i j} 
			+ 
			\delta G^{i j}(\vex{r})
		\right]
		T_{i j}
	+
	\e
	\left(\bar{R} L + \bar{L} R\right)
	\right\},
\end{align}
where repeated indices are summed over $i,j \in \{+,-\}$, 
and 
where the free-fermion stress tensor is 
\begin{align}
	T_{i j}
	=&\,
	-\frac{1}{2 \pi}
	\begin{bmatrix}
	T(z) & -2 \pi \, T_{+ -} \\
	-2 \pi \, T_{-+} & \bar{T}(\bar{z})
	\end{bmatrix}
\nonumber\\
	=&\,
	\frac{i}{2}
	\begin{bmatrix}
	\phantom{-} \bar{L} \!\stackrel{\leftrightarrow}{\parr}\! L 	& - \bar{R} \!\stackrel{\leftrightarrow}{\parr}\! R \\
	- \bar{L} \!\stackrel{\leftrightarrow}{\parb}\! L 		& \phantom{-} \bar{R} \!\stackrel{\leftrightarrow}{\parb}\! R
	\end{bmatrix},
\end{align}
with $z = x + i y$ and $\parr = (1/2)(\parr_x - i \parr_y)$.
The Euclidean metric in conformal coordinates is 
\[
	\eta^{i j} 
	\rightarrow 
	2 
	\begin{bmatrix}
	0 & 1 \\	
	1 & 0 
	\end{bmatrix}.
\]
Eq.~(\ref{SCFT}) shows that the effects of disorder couple gravitationally via
the metric perturbation $\delta G^{i j}$, while nonzero energy $\e$ 
couples to an (imaginary or tachyonic) mass operator for the fermions. 
The mass must be imaginary in order to ensure plane-wave correlations in the clean limit. 
The components of the perturbing 2D metric are 
\begin{align}
\begin{aligned}
	\delta G^{+ -} 
	=&\,
	\left(\dvoo + \dvtt + i \vot - i \vto\right),
\\
	\delta G^{- +} 
	=&\,
	\left(\dvoo + \dvtt - i \vot + i \vto\right),
\\
	\delta G^{+ +} 
	=&\,
	\left(- \dvoo + \dvtt - i \vot - i \vto\right),
\\
	\delta G^{- -} 
	=&\,
	\left(- \dvoo + \dvtt + i \vot + i \vto\right).
\end{aligned}
\end{align}

The five variants of the model (a)--(e) were defined and described below Eq.~(\ref{DiracRVH}). 
In the conformal language [Eq.~(\ref{SCFT})], models (c) and (d) are special.
Model (c) has $\delta G^{++} = \delta G^{--} = 0$; in this case, disorder couples
only to the ``diagonal'' components of the stress tensor $\{T_{+-},T_{-+}\}$. 
These are operators with conformal spin $s = 0$; the disorder induces isotropic 
flattening or steepening of the Dirac cone, as well as local rotations of the Dirac 
pseudospin $\{\sigh^{1,2}\}$ relative to the coordinate axes $\{x_{1,2}\}$. 
By contrast, model (d) has $\delta G^{+-} = \delta G^{-+} = 0$. 
In this case disorder couples only to the purely chiral $T(z)$ and $\bar{T}(\bar{z})$ fields,
which carry conformal spins $s = \pm 2$. These ``$d$-wave'' operators describe nematic deformations
of the Dirac cone. We will see that the nematic-only model (d) shows very robust ``critical stacking'' behavior,
while stacking in model (c) is disrupted by increasing the disorder strength.


\begin{figure}[b!]
\centering
\includegraphics[width=0.47\textwidth]{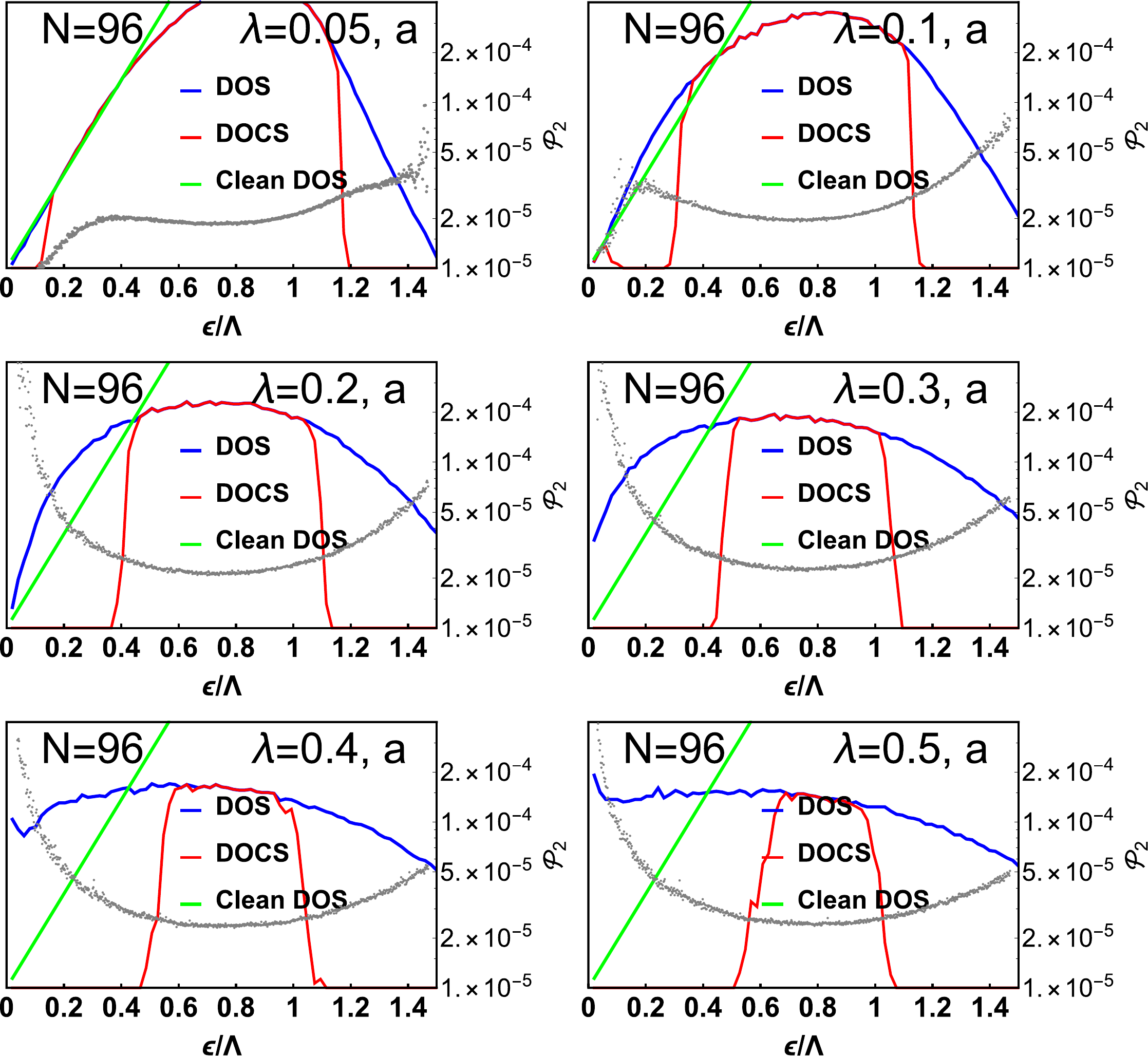}
\caption{This figure shows the 
total density of states (DOS) 
(blue)
and 
the 
density of critical states (DOCS)
(red)
for the Dirac model with QGD. 
Results are shown for model (a), defined below Eq.~(\ref{DiracRVH});
the clean DOS 
(green)
is also shown for comparison. 
Results obtain by diagonalizing the Hamiltonian in Eq.~(\ref{hMomForm}) over a 
$(2N+1)\times(2N+1)$ grid in momentum space, with $N = 96$ here. 
Data is plotted for the six different indicated values of the dimensionless disorder strength $\lambda$;
strong disorder corresponds to $\lambda \gtrsim 0.393$. 
The DOCS counts the number of states with critical statistics (multifractal spectra) 
that match a universal ansatz with a certain fitness criterion (see text).
Also plotted is the second IPR $\mathcal{P}_2$ (grey dots), defined by Eq.~(\ref{IPRDef}). 
For model (a), a large swath of the spectrum appears critical for weak disorder. 
However, as the disorder strength is increased, the swath shrinks. 
The IPR $\mathcal{P}_2$ shows that states outside of the swath are more rarified or localized
than the critical ones. 
The linear-in-energy DOS of the clean limit is strongly distorted and filled-in at low energies 
for $\lambda \gtrsim 0.2$. 
}
\label{Fig--DOS(a)}
\end{figure}

\begin{figure}[b!]
\centering
\includegraphics[width=0.47\textwidth]{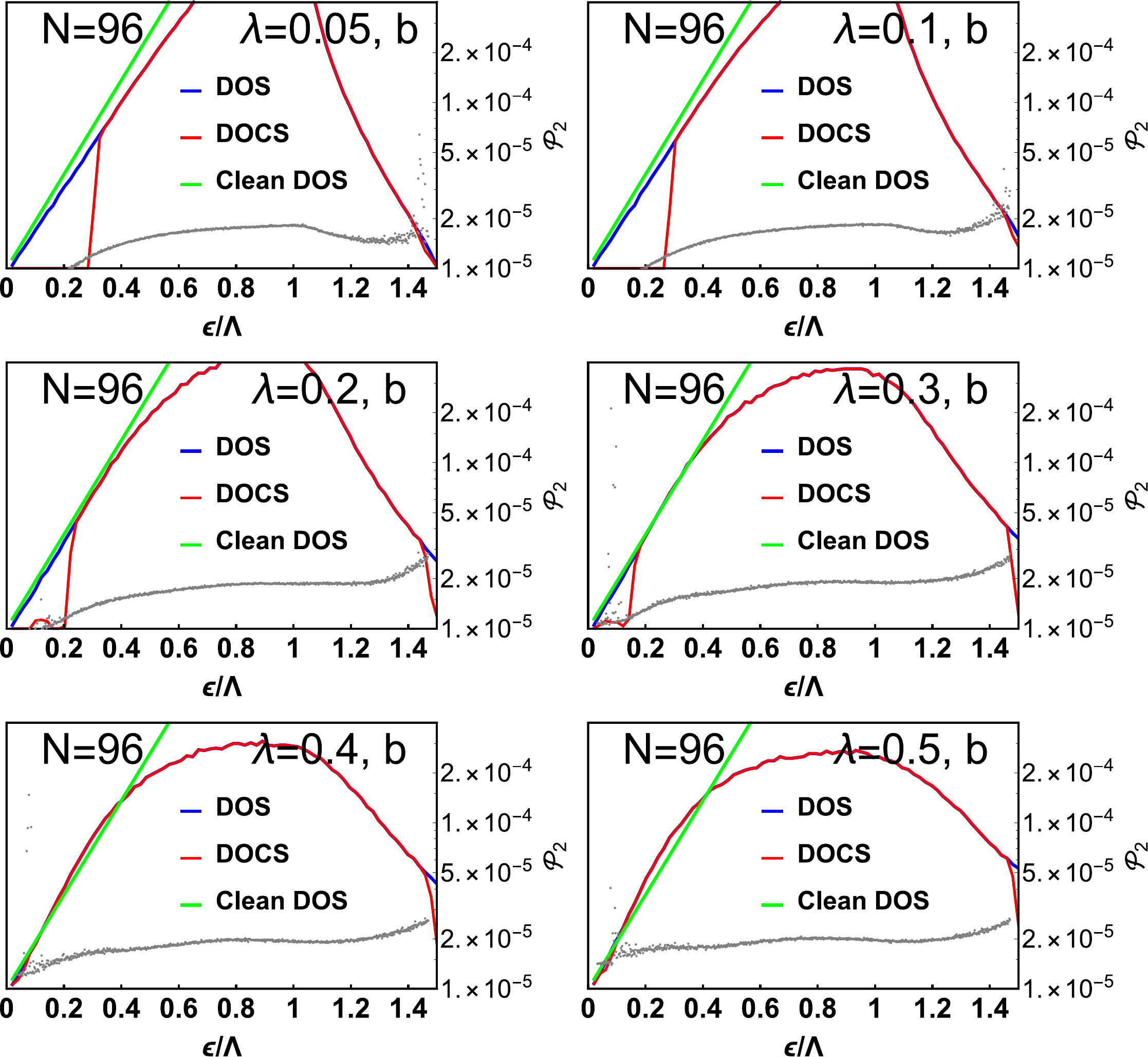}
\caption{The same as Fig.~\ref{Fig--DOS(a)}, but now for model (b).
In this case, the swath of critical states \emph{increases} with increasing disorder strengths. 
For $\lambda \lesssim 0.3$, the states at low energies are more plane-wave-like
(less rarified) than the critical ones, as indicated by the IPR $\mathcal{P}_2$;
some more rarified (supercritical) states appear at low energies for stronger disorder.
``Strong disorder'' corresponds to the threshold $\lambda \gtrsim 0.393$ (see text). 
With plane-wave-like states at low energy and a robust swath of strongly inhomogeneous, but extended
critical states at finite energy, model (b) exhibits a phenomenology most similar to STM 
data taken in the cuprate superconductor BSCCO \cite{Davis08,DavisReview}.
Note that the linear-in-energy low-energy total DOS persists to the largest disorder strength.  
As described below Eq.~(\ref{DiracRVH}), model (b) incorporates random rotations of the pseudospin
axes relative to the spatial coordinates, as well as nematic fluctuations of the Dirac cone. 
It \emph{excludes} isotropic flattening or steepening of the cone, as does model (d) (Fig.~\ref{Fig--DOS(d)}).
Corresponding position-space LDOS maps for model (b) are shown in Fig.~\ref{Fig--Maps(b)}.  
Representative multifractal spectra from the critical swath are depicted in Fig.~\ref{Fig--Dirt_Series_Dq(b)}. 
}
\label{Fig--DOS(b)}
\end{figure}

\begin{figure}[b!]
\centering
\includegraphics[width=0.47\textwidth]{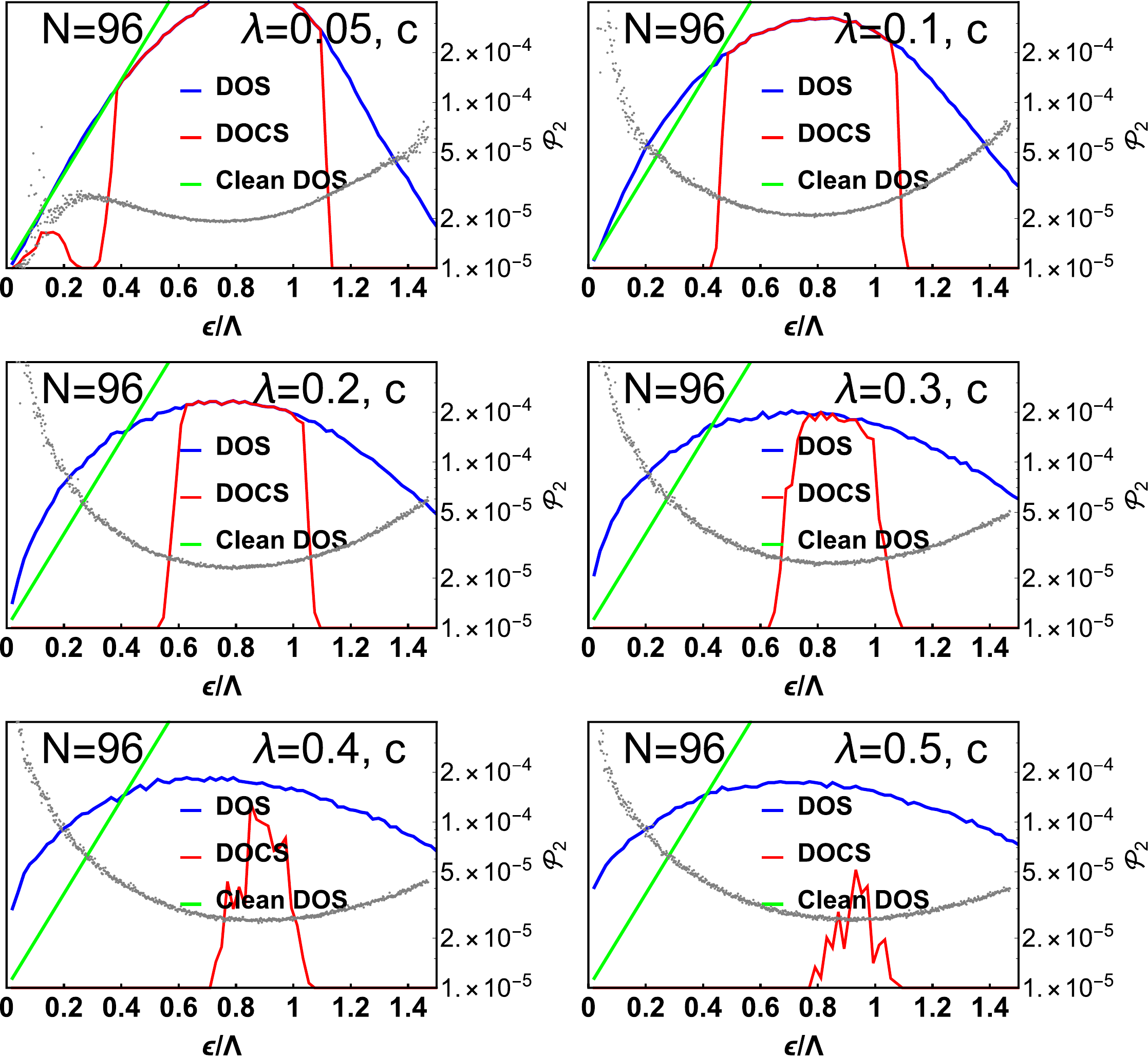}
\caption{The same as Fig.~\ref{Fig--DOS(a)}, but now for model (c).
In this case the critical swath decreases rapidly with increasing disorder strength. 
Both models (a) and (c) include isotropic flattening or steepening of the cone, which appears
to produce strong disorder effects. The states outside the critical swath are more affected
by the QGD (more rarified) than the critical ones, 
as indicated by the enhanced second IPR $\mathcal{P}_2$.
}
\label{Fig--DOS(c)}
\end{figure}

\begin{figure}[b!]
\centering
\includegraphics[width=0.47\textwidth]{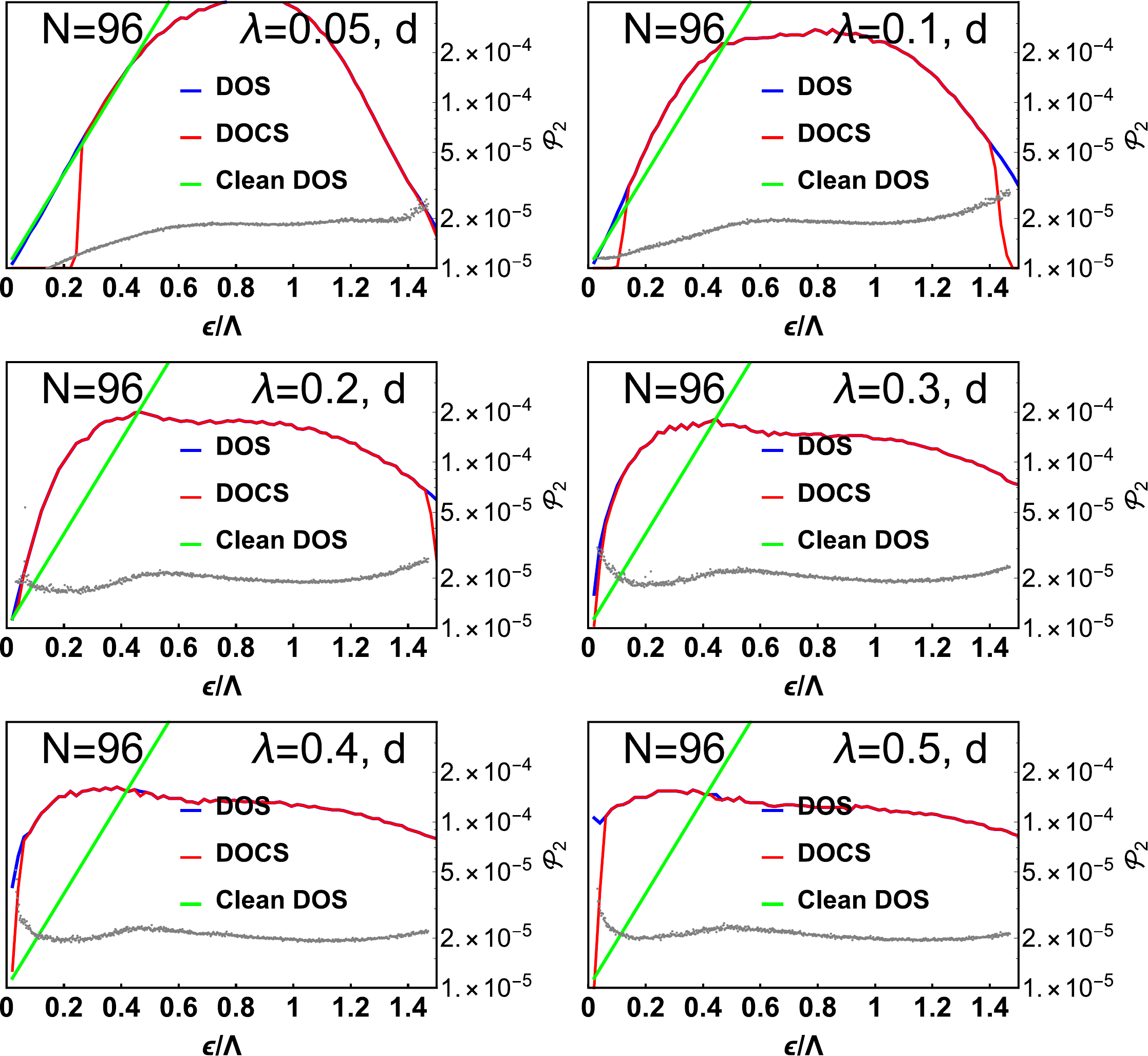}
\caption{The same as Fig.~\ref{Fig--DOS(a)}, but now for model (d).
Model (d) is similar to model (b), except that it excludes the random rotations incorporated in the latter. 
Both models (b) and (d) feature nematic randomness, i.e.\ random squishing of the Dirac cone. 
Similar to the data for model (b) shown in Fig.~\ref{Fig--DOS(b)}, 
the swath of critical states for model (d) \emph{increases} with increasing disorder strengths. 
Different from model (b), strong disorder effects occur at intermediate values of $\lambda$
near zero energy: the DOS fills in and flattens elsewhere, 
and $\mathcal{P}_2$ is enhanced for the lowest-energy states. 
Representative multifractal spectra from the critical swath are depicted in Fig.~\ref{Fig--Dirt_Series_Dq(d)}. 
}
\label{Fig--DOS(d)}
\end{figure}

\begin{figure}[b!]
\centering
\includegraphics[width=0.47\textwidth]{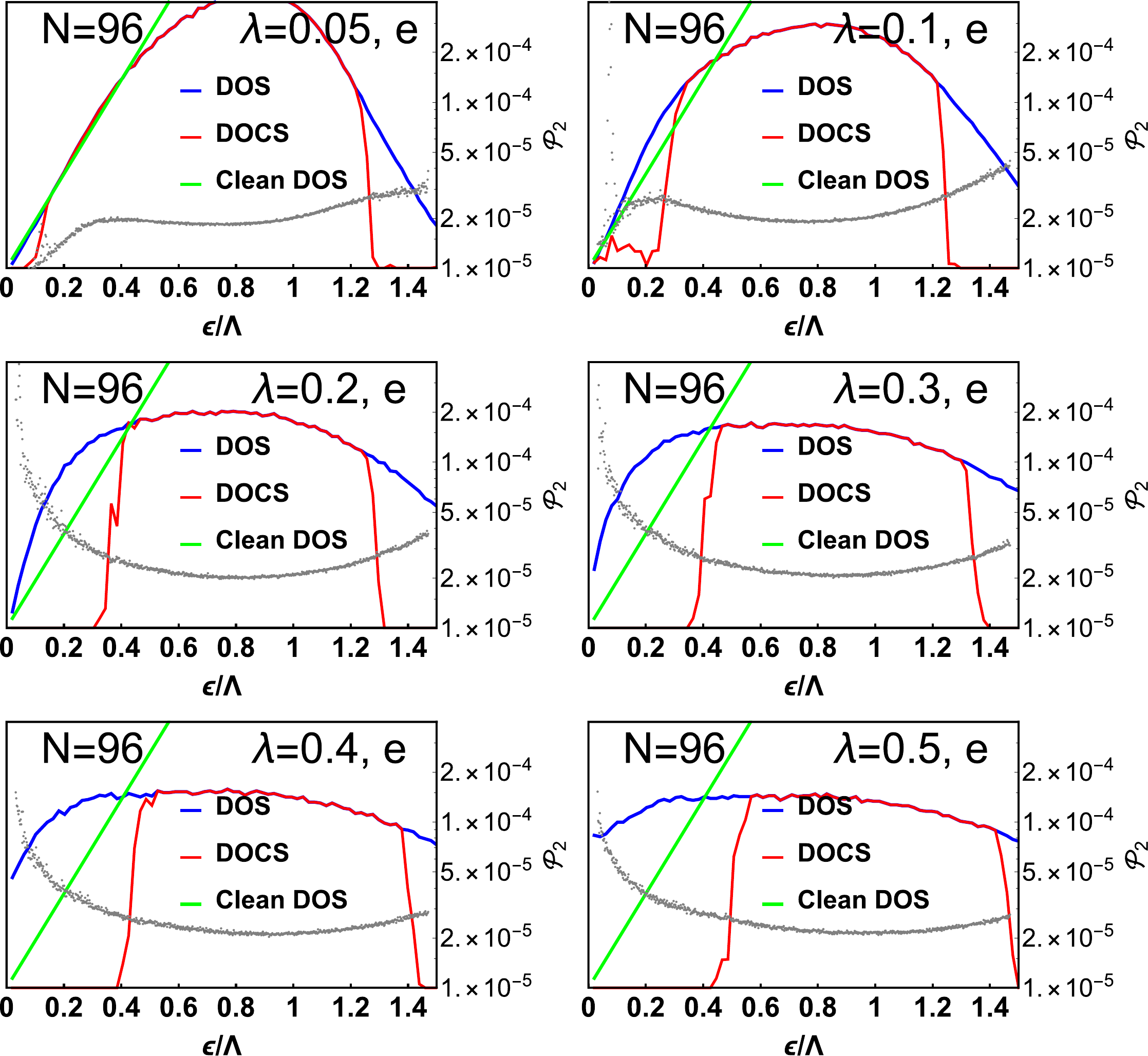}
\caption{The same as Fig.~\ref{Fig--DOS(a)}, but now for the generic model (e).
This model incorporates isotropic flattening and steepening of the Dirac cone, random rotations,
and nematic disorder. The behavior is intermediate between models (a),(c) and (b),(d). 
Like models (b) and (d), a wide swath of critical states persists for all disorder strengths. 
Like models (a) and (c), however, low-energy states become strongly affected by increasing randomness,
exhibiting supercritical (more rarified) behavior as indicated by the enhanced $\mathcal{P}_2$.
Corresponding position-space LDOS maps for model (e) are shown in Fig.~\ref{Fig--Maps(e)}.  
Representative multifractal spectra from the critical swath are depicted in Fig.~\ref{Fig--Dirt_Series_Dq(e)}. 
}
\label{Fig--DOS(e)}
\end{figure}


\begin{figure}[t!]
\centering
\includegraphics[width=0.45\textwidth]{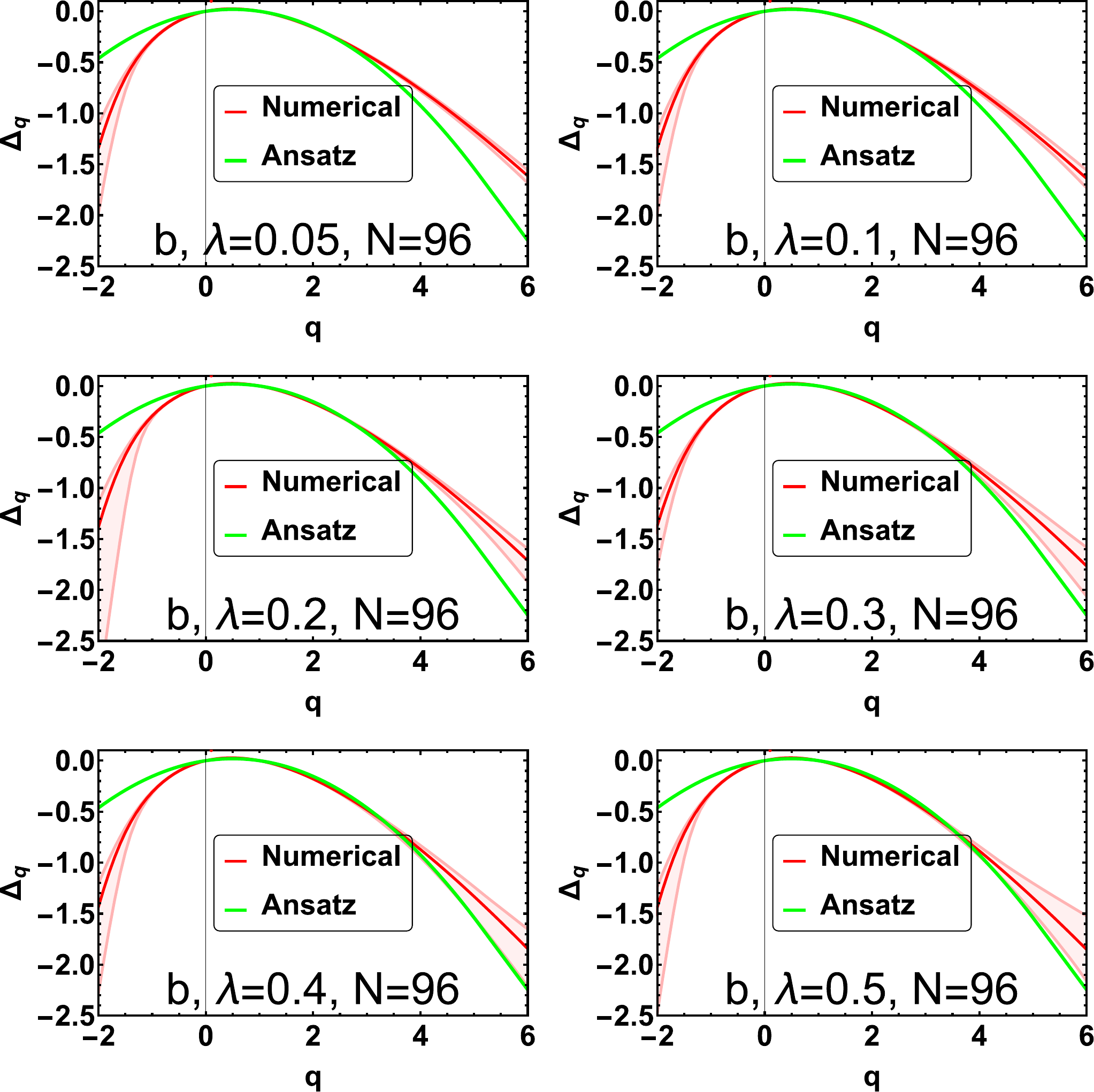}
\caption{Anomalous multifractal spectrum $\Delta(q) \equiv \tau(q) - 2 (q - 1)$
for an energy bin of states selected from the DOS with the highest percentage of 
critical states. Here the spectrum is shown for model (b), 
evaluated for the six different disorder strengths associated to the DOCS and DOS plots
in Fig.~\ref{Fig--DOS(b)}.
The solid red curve denotes an average over the 15 states in the bin;
the shaded red region indicates the standard deviation. 
The green curve is the parabolic ansatz for $\Delta(q)$, 
Eq.~(\ref{CI-MFC}) with $\theta = 1/13$. 
States contributing to the critical count (DOCS) in Fig.~\ref{Fig--DOS(b)}
match the parabolic ansatz within a certain thresold (see text) 
over the range $0 < q \leq q_c = 5.1$.}
\label{Fig--Dirt_Series_Dq(b)}
\end{figure}

QGD can affect massless 2+1-D Dirac carriers whenever the latter arise from a correlation gap. 
Strong spatial inhomogeneity has been observed in gap maps of the superconducting and 
pseudogap regimes in the $d$-wave cuprate superconductor BSCCO
\cite{Davis01,Davis02,Davis05-a,Davis05-b,Davis08,DavisReview}. 
This should in turn imply the modulation of the nodal quasiparticle 
velocities along the Fermi surface [see Fig.~\ref{Fig--RVC}(c)]. 

Additional sources of internode scattering in the cuprates 
can arise due to short-ranged impurities such as interstitial oxygen dopants 
\cite{DavisReview,AltlandSimonsZirnbauer02,Hirschfeld02,LeeMott06};
these can dominate over the effects of velocity modulation.
Depending on whether one, two, or four nodes are coupled in the $d$-wave problem,
one can get critical states all the way down to zero energy \cite{Ostrovsky2007,Chou2014,Ghorashi18}
(with a concomitant sublinear scaling of the low-energy global density of states) 
\cite{Ludwig1994,Nersesyan1994,Mudry1996,Caux1996,Bhaseen2001}.
The other possibility is Anderson localization across the energy spectrum for the most 
generic model of impurity scattering \cite{Senthil98,AltlandSimonsZirnbauer02};
see Sec.~\ref{Sec:DescModels} and Table~\ref{TSCTable} for a review of these models and 
scenarios. 
We only emphasize two points here. 
First, the 
linear low-energy density of states and the coherent, plane-wave-like
nature of the lowest-energy quasiparticles observed in STM studies \cite{Davis05-a,Davis05-b,Davis08,DavisReview}
appear inconsistent with strong internode scattering.
Second, the experimentally-observed thermal conductivity \cite{Taillefer00,Orenstein00} is not consistent with localization;
in fact, the ``universal result'' in Eq.~(\ref{TC}) is predicted to hold for any \emph{topologically restricted}
model of scattering, including Eq.~(\ref{DiracRV}) (see Sec.~\ref{Sec:DescModels}). 

Eq.~(\ref{DiracRV}) could be also realized on the surface of a class DIII
topological superconductor (TSC) 
\cite{TSCRev1,TSCRev2,TSCRev3,TSCRev4}
with winding number $|\nu| = 1$ \cite{Nakai14,Roy19}. 
In this case $\psi$ is a real Majorana field with $\bar{\psi}_\sigma = i \psi_{\sigma'} \left(\sigh^1\right)_{\sigma'\sigma}$, 
and one can show that (see Appendix \ref{Sec:3HeB})
\begin{align}\label{vDIII}
\begin{gathered}
	v_{11}(\vex{r}) 
	= 
	v_{22}(\vex{r})
	=
	v_0
	\left\{
	1
	+
	\vartheta
	\left[e A^0(\vex{r})/\ebulk\right]
	\right\},
\\
	v_{12} = v_{21} = 0,
\end{gathered}
\end{align}
where $v_0$ is the bare surface Majorana velocity,
$A^0(\vex{r})$ is the screened electric potential due to (e.g.) static charged impurities,
$\vartheta$ is a constant, and $\ebulk$ is the bulk gap energy of the TSC. 
The disorder can be characterized by a variance
$
	\tilde{\lambda}
	\propto
	\nimp \left[e^2 / \left(k_F \ktf\right)\right]^2,
$
where $\nimp$ is the surface areal impurity density, and
$k_F$ ($\ktf$) is the bulk Fermi (Thomas-Fermi screening) wave vector.
In units such that $v_0 = 1$, $\tilde{\lambda}$ is a squared-length; 		
weak, short-range correlated 
QGD
is therefore \emph{strongly irrelevant} \cite{KPZ--Note}
at zero
energy on the surface. 
Low-energy states are thus expected to be only weakly affected by QGD \cite{Nakai14}. 
The fate of the finite-energy states 
is not obvious, however, since energy is itself \emph{strongly relevant} \cite{Ludwig1994,Ghorashi18}.

\subsection{Density of critical states (DOCS) and disorder-strength scaling \label{Sec:NumResults}}

In order to work directly in the continuum,
we diagonalize Eq.~(\ref{DiracRVH}) exactly 
in momentum space,
	\begin{align}\label{hMomForm}
	\hat{h}_{\vex{k},\vex{k'}}
	=&\,
	\begin{bmatrix}
	0 & \left(k_x - i k_y\right)^\nu \\
	\left(k_x + i k_y\right)^\nu & 0 
	\end{bmatrix}
	\delta_{\vex{k},\vex{k'}}
\nonumber\\
&\,
	+
	\left(\frac{k_x + k_x'}{2}\right)
	\left[
		\sigh^1 
		\,
		\delta v_{11}
		+
		\sigh^2 
		\,
		v_{21}
	\right]\left(\vex{k} - \vex{k'}\right)
\nonumber\\
&\,
	+
	\left(\frac{k_y + k_y'}{2}\right)
	\left[
		\sigh^1 
		\,
		v_{12}
		+
		\sigh^2 
		\,
		\delta v_{22}
	\right]\left(\vex{k} - \vex{k'}\right).
\end{align}	
We set the parameter $\nu = 1$ in Eq.~(\ref{hMomForm}),
appropriate to the relativistic clean system. 
A higher odd-integral value of $\nu \in \{3,5,\ldots\}$ can 
be used to represent a 2D class DIII Majorana surface fluid
that arises from a bulk topological superconductor with
corresponding winding number $\nu$ \cite{Roy19}. We used Eq.~(\ref{hMomForm})  
to analyze the multifractal spectra of low-energy surface
states with $\nu \in \{3,5,7\}$ in Ref.~\cite{Roy19}, 
where we tested predictions of the class DIII SO($2n$)$_\nu$
($n \rightarrow 0$)  
conformal field theory expected to describe the zero-energy surface states
in these cases \cite{Foster14}.   

The impurity velocity potentials in Eq.~(\ref{hMomForm})
are each taken to be a composition of random phases in momentum space, e.g.\
\begin{align}\label{vParam}
	\delta v_{11}(\vex{k}) 
	= 
	\left({\sqrt{\tilde{\lambda}_{11}}}/{L}\right) 
	\exp\left[i \theta_{11}(\vex{k}) - k^2 \zeta^2/4\right].
\end{align}
Here $\theta_{11}(-\vex{k}) = - \theta_{11}(\vex{k})$; 
these are otherwise independent, uniformly distributed phase angles. 
We choose a short correlation length so as to approximate white noise disorder, appropriate (e.g.) 
to model the nanoscale gap inhomogeneity observed in the cuprates \cite{DavisReview}, or
efficient screening of Coulomb impurities on the surface of a TSC: 
$\zeta \equiv (0.25) (2 \pi / \Lambda)$, where $\Lambda$ is the momentum cutoff. 
Disorder becomes ``strong'' when the local variance of the velocity components
in position space ($\equiv \Delta v_{i j}$) 
becomes of order one. Disorder beyond this threshold regularly tips the velocity components through zero,
which can create curvature horizons \cite{Volovik,Jafari19} [see Eq.~(\ref{RScalar})].
Since $\Delta v_{i j} = \sqrt{\tilde{\lambda}_{i j}/(2 \pi \zeta^2)}$, 
this corresponds to the condition $\lambda_{i j} \equiv \tilde{\lambda}_{i j} (\Lambda/2\pi)^2 = \pi/8 \simeq 0.393$.
Here $\lambda_{i j}$ denotes the dimensionless disorder strength. 

We study the five model variants (a)--(e) defined below Eq.~(\ref{DiracRVH}).
For simplicity, we assign the same dimensionless disorder variance $\lambda$ to all
independent, nonzero velocity potentials in each variant.

Representative plots of the LDOS for states at different energies 
in models (b) and (e)
appear in 
Figs.~\ref{Fig--Maps(b)} and \ref{Fig--Maps(e)}. 
Wave function quantum criticality is characterized by the spectrum of exponents $\tau(q)$, reviewed
in Sec.~\ref{Sec:CI-2} [Eqs.~(\ref{tau(q)Def}) and (\ref{CI-MFC})]. 
We calculate the multifractal spectrum $\tau(q)$ by the usual box-counting method; 
the reader is referred to Ref.~\cite{Ghorashi18} for technical details. 
In order to quantify the degree of criticality throughout
the energy spectrum, we employ the following criterion. 
We compare the computed $\tau(q)$ spectrum 
for every state in regularly spaced
energy bins to a quadratic ansatz \cite{Ghorashi18}, 
$\tau(q) =  2(q-1)(1 - q/q_c^2)$ for $|q| \leq q_c$. 
We employ the ``fitness'' criteria, 
defined as follows \cite{Ghorashi18}. 
For each eigenstate $\psi(\vex{r})$, we compute the error between the numerical spectrum 
[$\equiv \tau_N(q)$] and the appropriate analytical prediction [$\equiv \tau_A(q)$], 
error$(q) \equiv |\tau_N(q)-\tau_A(q)|/\tau_A(q)$. 
If the error is less than or equal to 4\% for 85\% 
of the evaluated $q$-points in the interval $0 \leq q \leq q_c$, we keep the state. 
We consider 200 total bins across the energy spectrum.
In each bin we analyze 15 states with equally spaced eigenenergies; the DOCS is
computed from the fraction of these that satisfy the fitness criterion. 

\begin{figure}[t!]
\centering
\includegraphics[width=0.45\textwidth]{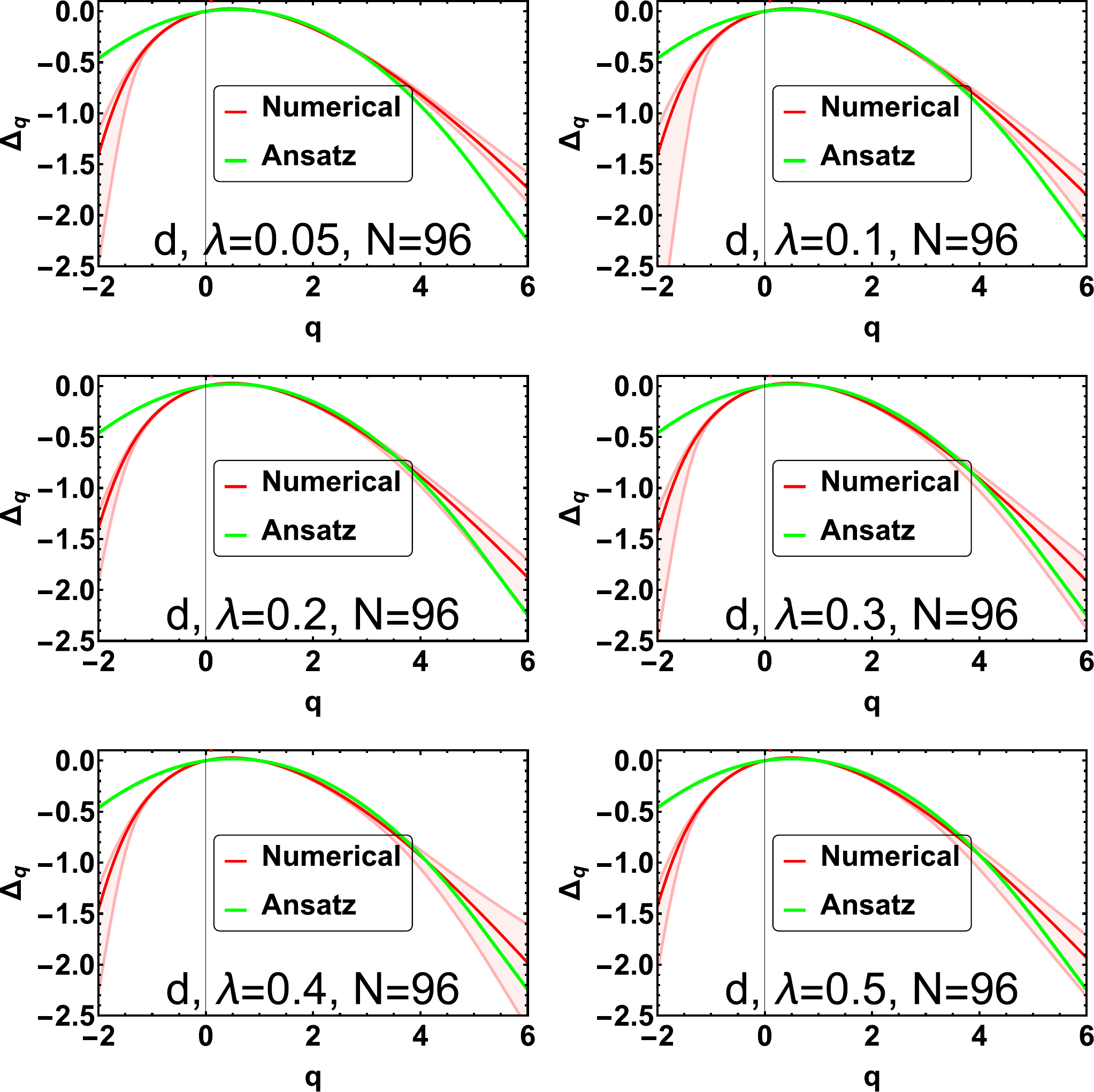}
\caption{The same as Fig.~\ref{Fig--Dirt_Series_Dq(b)}, but for model (d).
}
\label{Fig--Dirt_Series_Dq(d)}
\end{figure}

\begin{figure}[t!]
\centering
\includegraphics[width=0.45\textwidth]{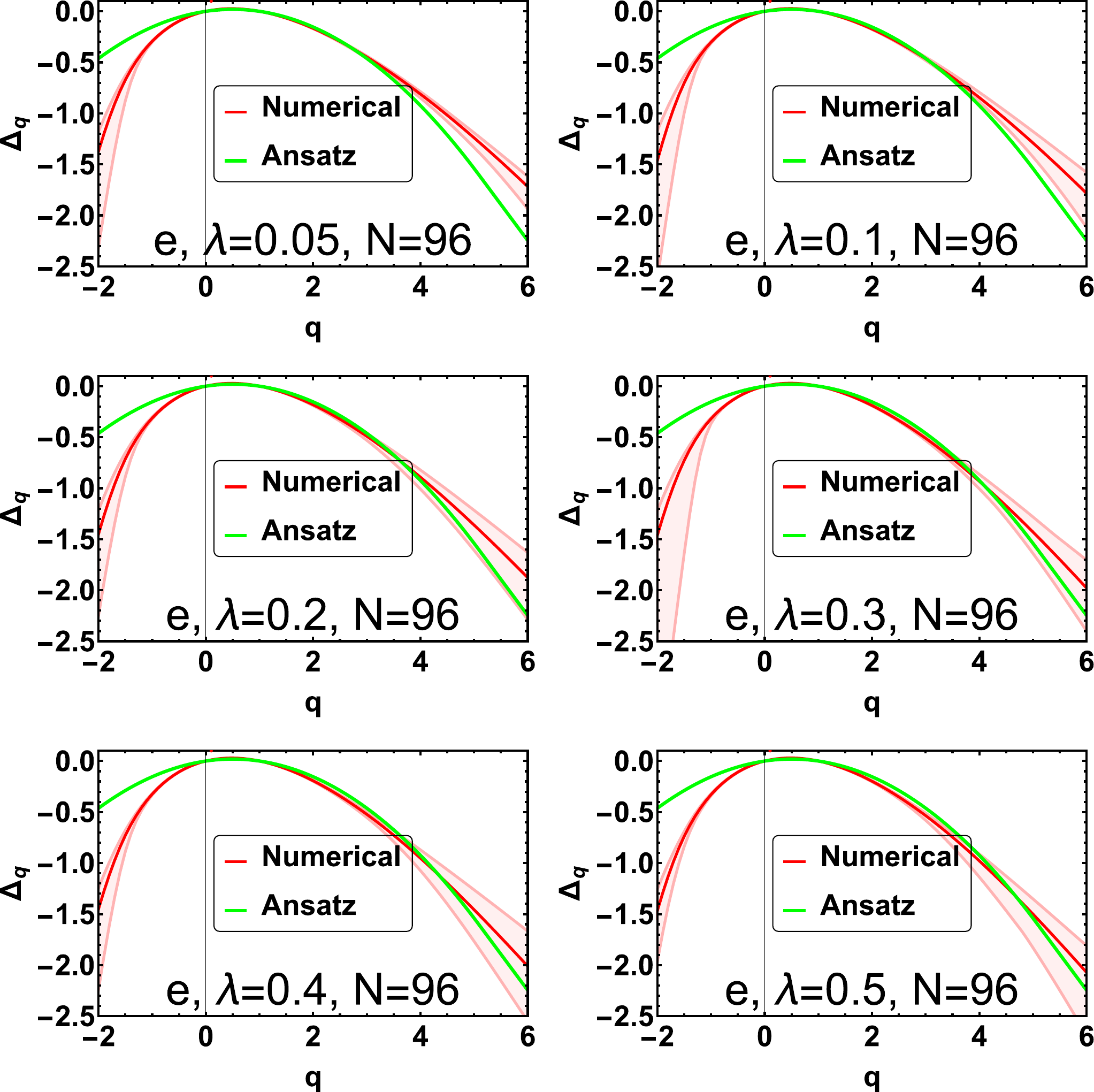}
\caption{The same as Fig.~\ref{Fig--Dirt_Series_Dq(b)}, but for model (e).
}
\label{Fig--Dirt_Series_Dq(e)}
\end{figure}

We empirically choose $q_c = 5.1$ for the parabolic ansatz;
this corresponds to $\theta \simeq 1/13$ in Eq.~(\ref{CI-MFC}).
Representative anomalous multifractal spectra 
$\Delta(q) \equiv \tau(q) - 2(q-1)$
are shown in 
Figs.~\ref{Fig--Dirt_Series_Dq(b)},
\ref{Fig--Dirt_Series_Dq(d)},
and
\ref{Fig--Dirt_Series_Dq(e)}
for the models (b), (d), and (e) that exhibit robust critical stacking. 
When we scan through states at different energies to evaluate
their ``fitness'' relative to the universal critical spectrum conjectured
above, we
exclude negative moments $q < 0$, since evaluating these
accurately requires significant coarse-graining; for this reason negative moments are typically not reported. 
We are unable to determine if the deviation seen between the ansatz
and the data for $q < 0$ in 
Figs.~\ref{Fig--Dirt_Series_Dq(b)}--\ref{Fig--Dirt_Series_Dq(e)} is intrinsic, or simply a finite-size limitation. 

We define the density of critical states (DOCS) as the number of states within an energy bin satisfying the above criterion. 
The ratio of the DOCS to the total density of states (DOS) is the effective energy-resolved distribution function for critical states \cite{Ghorashi18}.
For models (a)--(e), the DOS and the DOCS are shown in Figs.~\ref{Fig--DOS(a)}--\ref{Fig--DOS(e)}, respectively.
In each case, we show results for the maximum system size (193$\times$193 momentum grid) 
and six different disorder strengths $\lambda \in \{0.05,0.1,0.2,0.3,0.4,0.5\}$;
$\lambda \gtrsim 0.393$ corresponds to strong disorder (see above). 
In these plots, we also exhibit the bare second inverse participation ratio (IPR) 
\begin{align}\label{IPRDef}
	\mathcal{P}_2 \equiv \int d^2\vex{r} \, |\psi(\vex{r})|^4.
\end{align} 
This gives a raw measure of the spatial extent of each wave function. 
For a plane wave 
$\mathcal{P}_2 \sim 1/L^2$, 
while for a localized state with $\zeta_{\mathsf{loc}} \ll L$,
$\mathcal{P}_2 \sim 1/\zeta_{\mathsf{loc}}^2$, where $L$ and $\zeta_{\mathsf{loc}}$ denote the system size and localization length,
respectively. 
More generally, $(L^2) \mathcal{P}_2$ is enhanced for any rarification of the wave function (relative to the uniform plane wave case).


All five models exhibit a swath of critical states at finite energy for sufficiently weak disorder.
As we discuss below, we find very weak system-size dependence for these results. 
A crucial difference is how the critical swath varies as the strength of the disorder is increased. 
Figs.~\ref{Fig--DOS(a)} and \ref{Fig--DOS(c)} show that the swath shrinks for models (a) and (c) with increasing disorder; this is especially 
pronounced for model (c).
The second IPR $\mathcal{P}_2$ in these plots indicates that states outside of the critical swath are more rarified (supercritical),
except near zero energy for the weakest disorder strength.
As described below Eq.~(\ref{DiracRVH}), models (a) and (c) both feature disorder that locally flattens or steepens the Dirac cone 
in an isotropic fashion; model (a) also incorporates nematic fluctuations. 

By contrast, 
Figs.~\ref{Fig--DOS(b)} and \ref{Fig--DOS(d)}
show that the width of the critical swath for models (b) and (d) \emph{increases} with increasing disorder strength.
In model (b), in particular, the main effect of increasing the disorder is to lower the 
crossover energy ($\equiv \Delta_0$) between low-energy plane-wave-like states and finite-energy critical states.
The coexistence of low-energy plane-waves and finite-energy critical states (with fixed, universal multifractal fluctuations in the latter) 
over a wide range of disorder strengths means that model (b) is closest to the phenomenology of the cuprates, i.e.\ to the
energy-dependence of the LDOS maps observed in STM studies of BSCCO \cite{Davis08,DavisReview}.
Model (d) behaves in a similar fashion, except that the states near zero energy begin to show more rarified behavior 
(larger $\mathcal{P}_2$) at intermediate disorder strengths, even while the transition to ``stacked'' criticality 
decreases in energy. In model (d), this is concomitant with a filling in of the total DOS at zero energy. 
Models (b) and (d) feature \emph{nematic} disorder; model (d) corresponds to the pure ``$T\bar{T}$'' deformation 
in the CFT language of Eq.~(\ref{SCFT}) (after averaging over disorder, using e.g.\ replicas). 
Model (b) additionally incorporates local rotations. 

The DOS and DOCS for the generic model (e) are depicted in Figs.~\ref{Fig--DOS(e)}.
In this case the behavior is intermediate: a wide critical swath appears for all disorder strengths, but is 
pushed to higher and higher energies with increasing $\lambda$. Except for the weakest disorder, states
near zero energy become more rarified than those in the critical swath, as indicated by the enhancement of $\mathcal{P}_2$.

The anomalous multifractal spectra selected from an energy bin where the ratio of the DOCS to DOS is maximized
for models (b), (d), and (e)
are plotted for the same six disorder strengths in Figs.~\ref{Fig--Dirt_Series_Dq(b)}--\ref{Fig--Dirt_Series_Dq(e)}.
Note that despite the variation of $\lambda$ by an order of magnitude, there is little change in the
spectra. This is completely different from the conventional symmetry-based prediction reviewed in Appendix~\ref{Sec:Sym},
which argues that finite-energy states of the class DIII model in Eq.~(\ref{DiracRVH}) should reside in the 
symplectic class \cite{Evers2008}. In that case, although multifractality would be expected for weakly antilocalizing
states in finite volume, the curvature of the spectrum is predicted to scale linearly in the disorder strength 
or inverse bare (semiclassical) conductance [see Eqs.~(\ref{AII_Delta}) and (\ref{AII_G(L)}), below].


\begin{figure}[b!]
\centering
\includegraphics[width=0.3\textwidth]{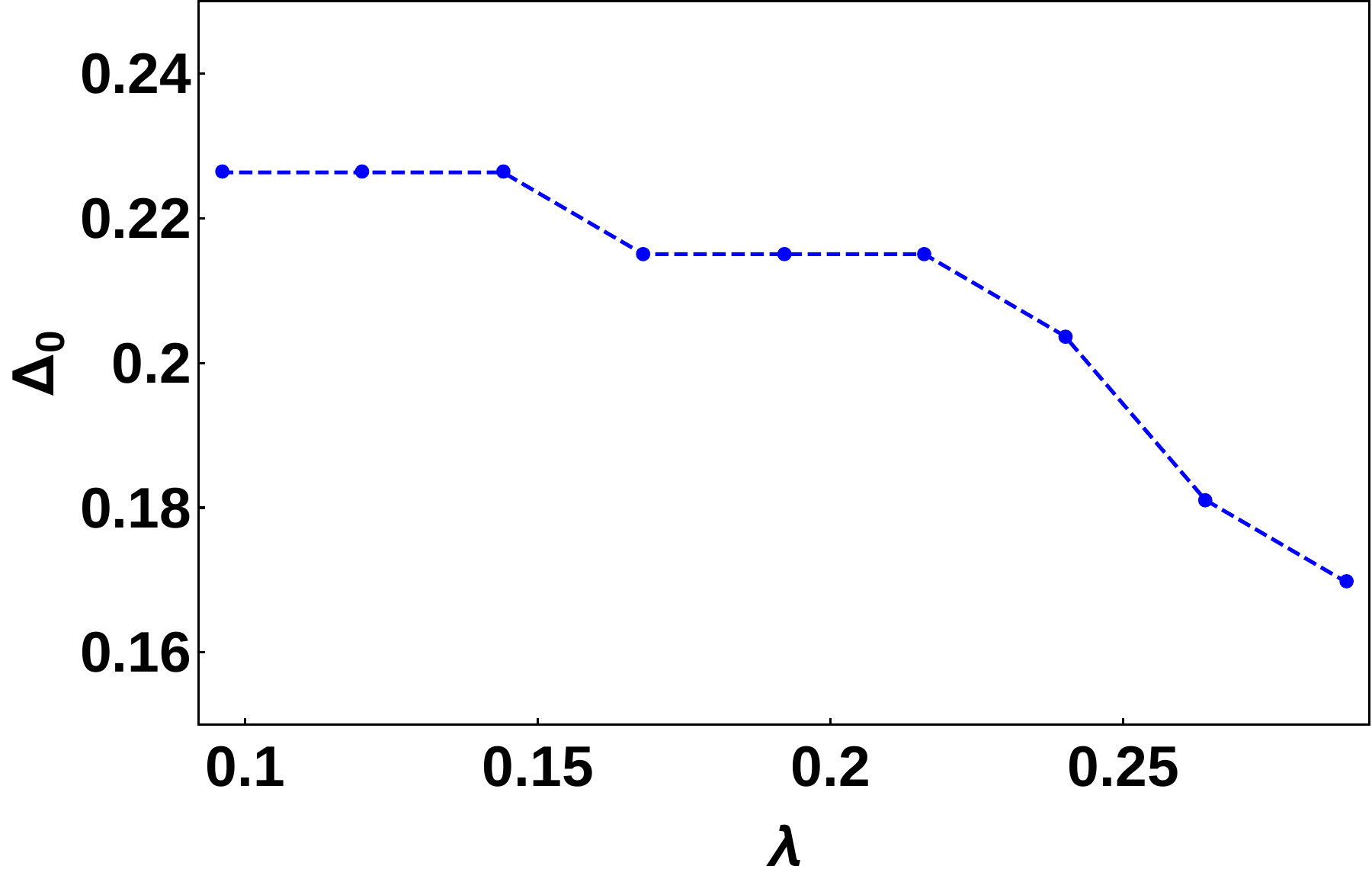}
\caption{Plot of the effective energy scale $\Delta_0$ versus disorder strength $\lambda$,
wherein the low-energy wave functions transition from plane-wave-like
to critical behavior. I.e., $\Delta_0$ is the energy scale measured from the Dirac point 
(in units of the cutoff $\Lambda$) 
at which the DOCS first becomes appreciable [Eq.~(\ref{Delta0Def})]. 
Here results are averaged over 10 disorder realizations for $N = 36$. 
}
\label{Fig--Delta0vsDirt}
\end{figure}


\begin{figure}[t!]
\centering
\includegraphics[width=0.47\textwidth]{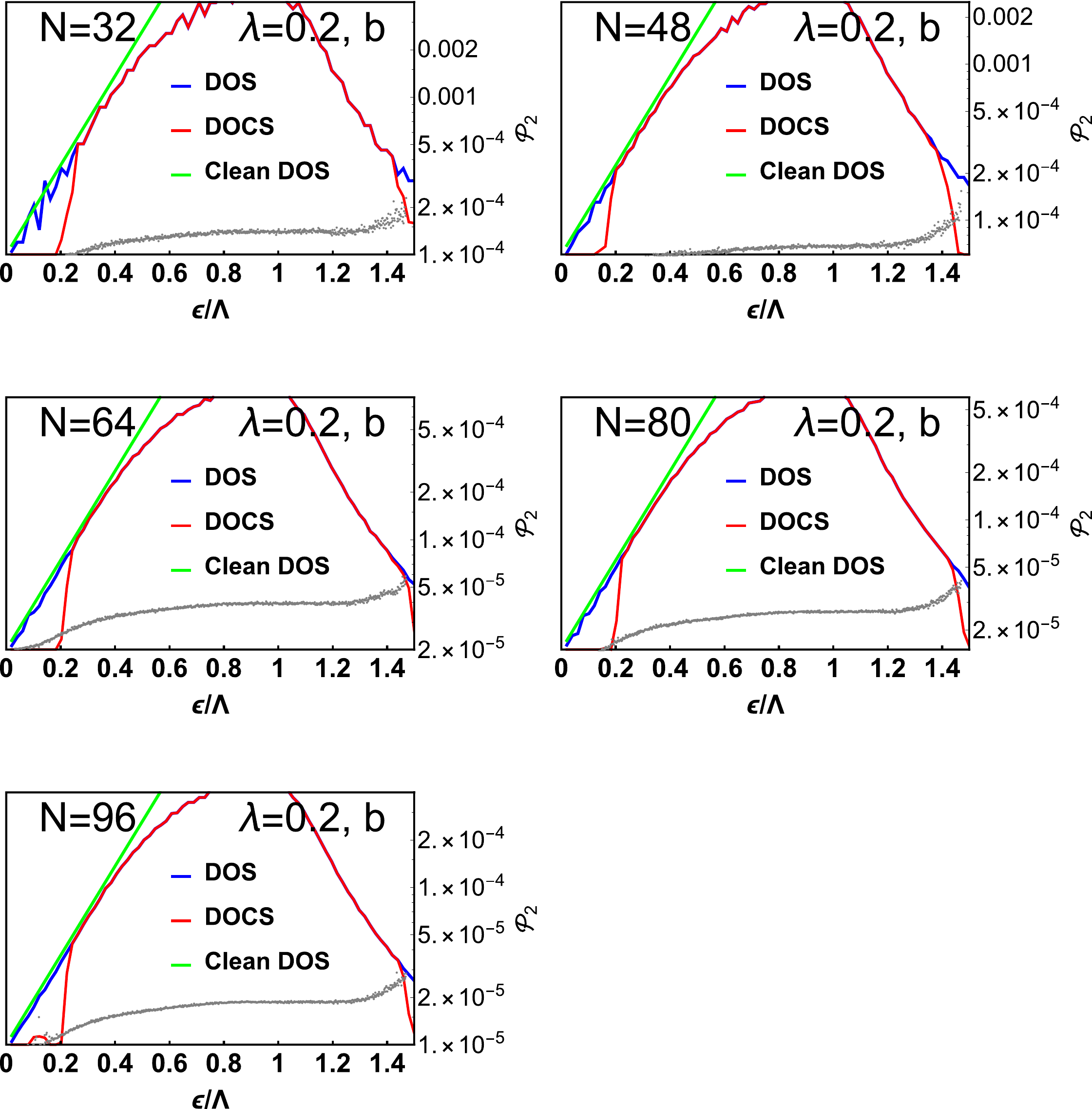}
\caption{Total DOS and DOCS for model (b) as in Fig.~\ref{Fig--DOS(b)}, 
but for five different system sizes. 
The dimensionless disorder strength is $\lambda = 0.20$; 
strong disorder corresponds to $\lambda \simeq 0.393$. 
}
\label{Fig--NSeriesDOCS(b)}
\end{figure}


\subsection{Plane-wave to critical-state crossover energy ``$\Delta_0$'' versus disorder strength}

A key observation in STM studies of BSCCO in the optimal to underdoped regime 
is the qualitative separation between
local density of states (LDOS) spectra obtained below and above a weakly doping-dependent
scale $\Delta_0$. At energies below (above) $\Delta_0$, the LDOS maps exhibit 
robust energy-dispersing quasiparticle interference (strong energy-independent, nanometer-scale spatial inhomogeneity),
interpreted as separating ``coherent'' low-energy quasiparticle states from ``incoherent'' intermediate-energy 
excitations \cite{Davis08,DavisReview}. 

For the QGD model (b) [defined below Eq.~(\ref{DiracRVH})] that 
best fits the cuprate phenomenology (see Figs.~\ref{Fig--Maps(b)} and \ref{Fig--DOS(b)}), 
we can try to compute the scale $\Delta_0$ in terms of the energy cutoff
$\Lambda$, as a function of the dimensionless disorder strength $\lambda$. We define $\Delta_0$ as 
the threshold where the DOCS becomes larger than a certain percentage of the total DOS. 
We choose the (arbitrary) criterion 
\begin{align}\label{Delta0Def}
	\text{DOCS} = (10 \, \%) \; \text{DOS}
\end{align}
in order to define $\Delta_0$, measured relative to the Dirac point. 
Results are shown in Fig.~\ref{Fig--Delta0vsDirt}. We find a weak decrease of $\Delta_0$ with increasing 
disorder strength.
This is qualitatively similar to cuprate data \cite{Davis08,DavisReview} with increased underdoping,
if the magnitude of the fluctuations in $\Delta_1(\vex{r})$ 
is taken as a proxy for the effective disorder strength (instead of the doping level).


\begin{figure}[t!]
\centering
\includegraphics[width=0.47\textwidth]{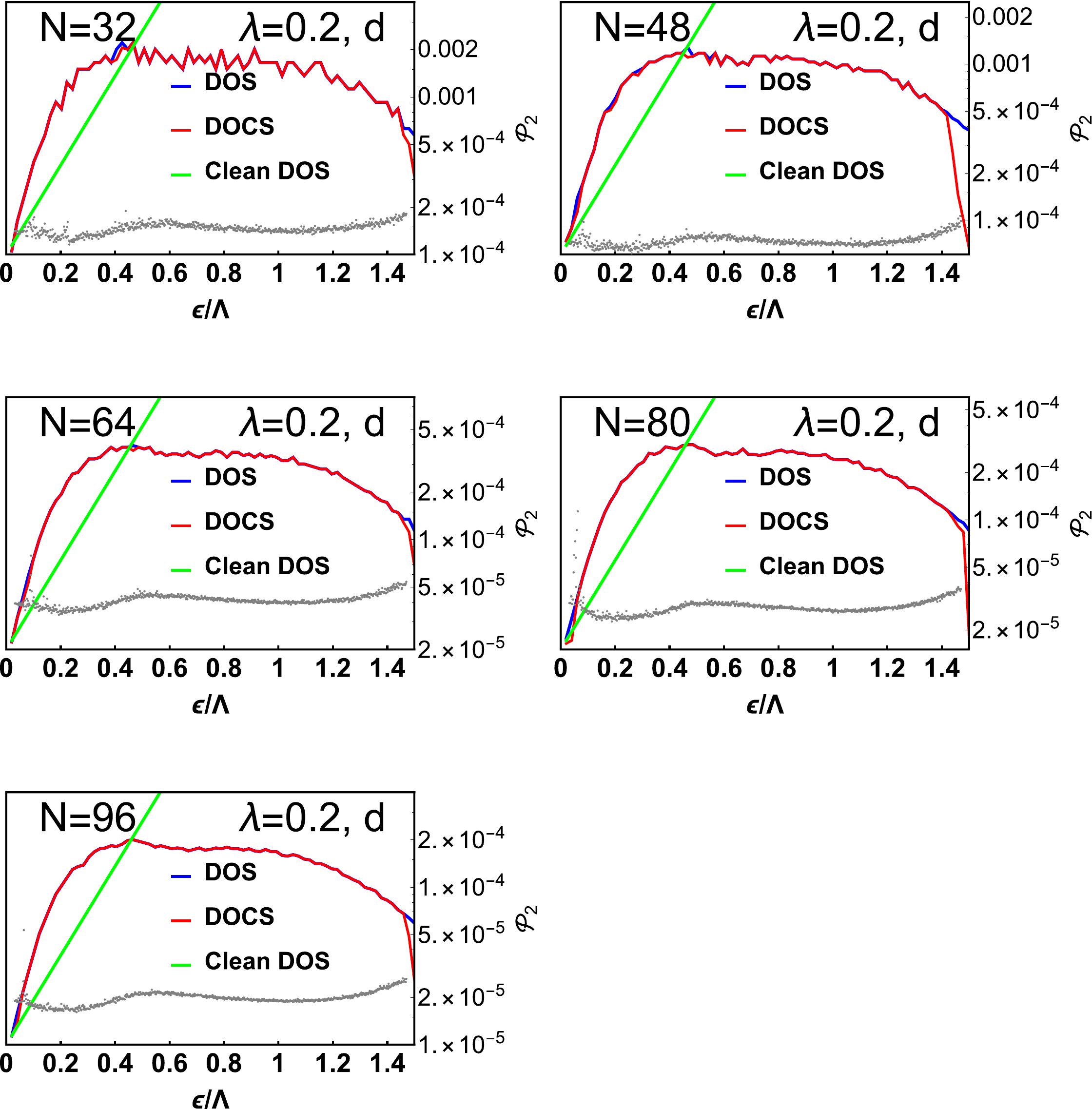}
\caption{Total DOS and DOCS for model (d) as in Fig.~\ref{Fig--DOS(d)}, 
but for five different system sizes. 
}
\label{Fig--NSeriesDOCS(d)}
\end{figure}


For the class DIII model (b), 
we are unable to extract a clear trend of $\Delta_0$ with respect
to the available system sizes $N$ (see Fig.~\ref{Fig--NSeriesDOCS(b)}).
In the other studied instances of quantum-critical wave-function stacking, 
the disorder is of vector-potential type \cite{Foster14}
(see Table~\ref{TSCTable} in Sec.~\ref{Sec:Review}). 
For short-range correlated vector-potential randomness, the disorder strength is intrinsically dimensionless. 
This leads to the expectation that the crossover energy scale analogous to $\Delta_0$ must
go to zero as the system size $N \rightarrow \infty$, consistent with the numerical
results obtained for class CI and AIII surface states in Refs.~\cite{Ghorashi18,Sbierski}.
For QGD studied here, however, the intrinsic disorder strength $\tilde{\lambda}$
has units of length squared. Then we can write
\begin{align}\label{Delta_0-Scaling}
	\Delta_0 = \left(1 / \sqrt{\tilde{\lambda}}\right) F\left({\tilde{\lambda}} / {L^2}\right),
\end{align}
where $L$ is the dimensionful linear system size,
and we have assumed 
that the scaling function $F$ does not depend explicitly on the cutoff $\Lambda$
(as usual \cite{GoldenfeldBook}).
Clearly we must have $\Delta_0 \rightarrow \infty$ as $\tilde{\lambda} \rightarrow 0$. 
We cannot rule out that the limit $F(x \rightarrow 0) > 0$, which would imply 
a finite $\Delta_0 \sim \tilde{\lambda}^{-1/2}$ in the $L \rightarrow \infty$ limit. 
As evidenced by the roughness of the 
trend indicated in Fig.~\ref{Fig--Delta0vsDirt}, much more work is needed to 
establish the behavior of $\Delta_0$ with respect to disorder and system size.  

In the QGD stacking scenario, a finite $\Delta_0$ would seemingly imply a finite-energy
semimetal to quantum-critical stacked transition. A crucial goal for future work is to establish
whether this transition exists at $\e > 0$, and if so what independent detectable signatures 
could be associated to it. In experiments on BSCCO, $\Delta_0$ can also be extracted from a ``kink'' 
in the energy-dependence of the spatially averaged tunneling spectrum \cite{DavisReview}.
The precise variation of the DOS at energies away from $\e = 0$
is however likely nonuniversal and model-dependent \cite{Evers2008}, in contrast to the universal 
\emph{spatial} fluctuations of the critical states in the stacking scenario \cite{Ghorashi18,Sbierski}.
Finally, we note that a kink in the DOS observed in experiments at the threshold of critical stacking might arise due
to the interplay with quasiparticle-quasiparticle interactions (i.e., an Altshuler-Aronov-type correction) \cite{LeeRamakrishnan}.


\begin{figure}[t!]
\centering
\includegraphics[width=0.47\textwidth]{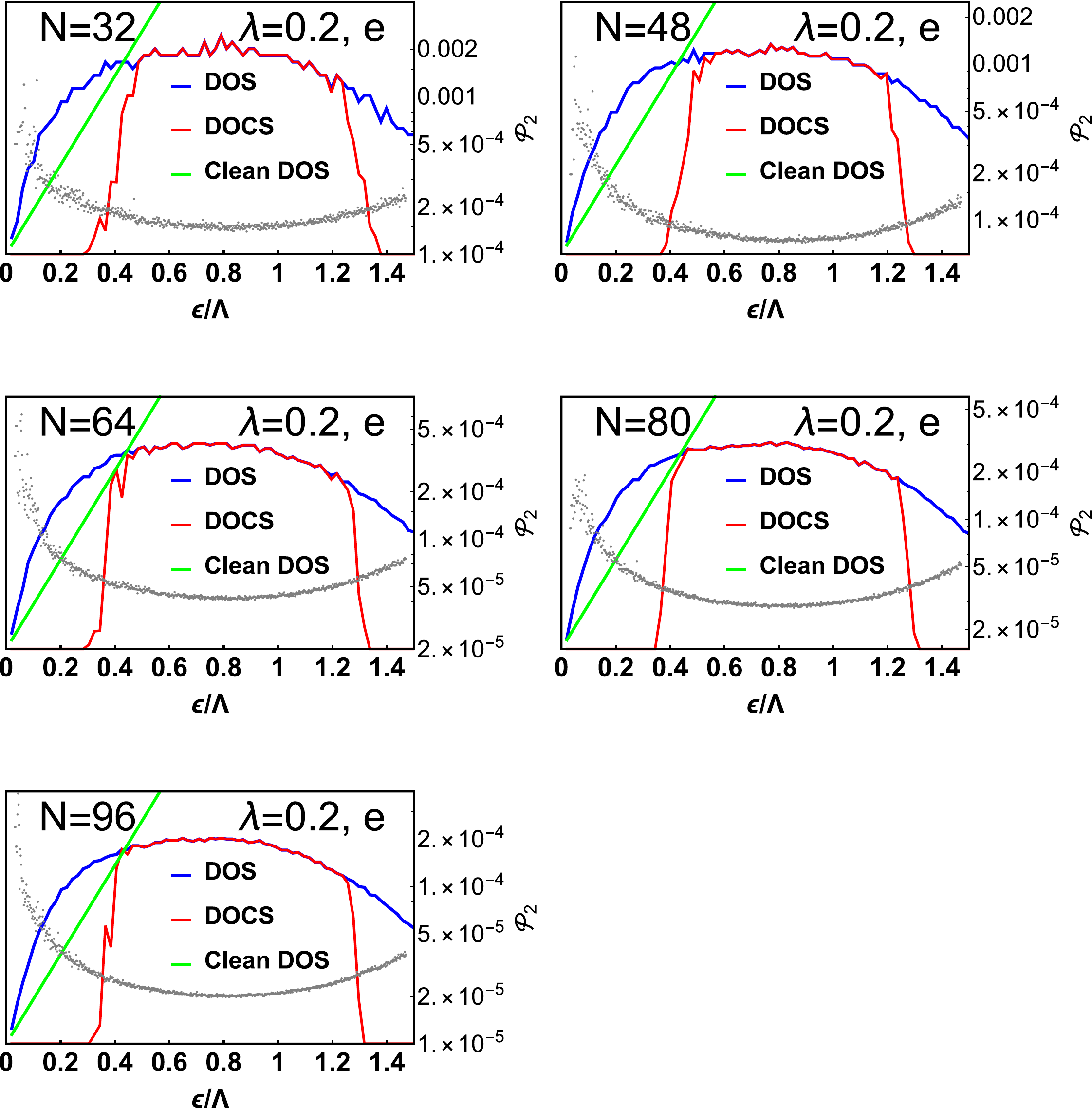}
\caption{Total DOS and DOCS for model (e) as in Fig.~\ref{Fig--DOS(e)}, 
but for five different system sizes. 
}
\label{Fig--NSeriesDOCS(e)}
\end{figure}


\begin{figure}[t!]
\centering
\includegraphics[width=0.47\textwidth]{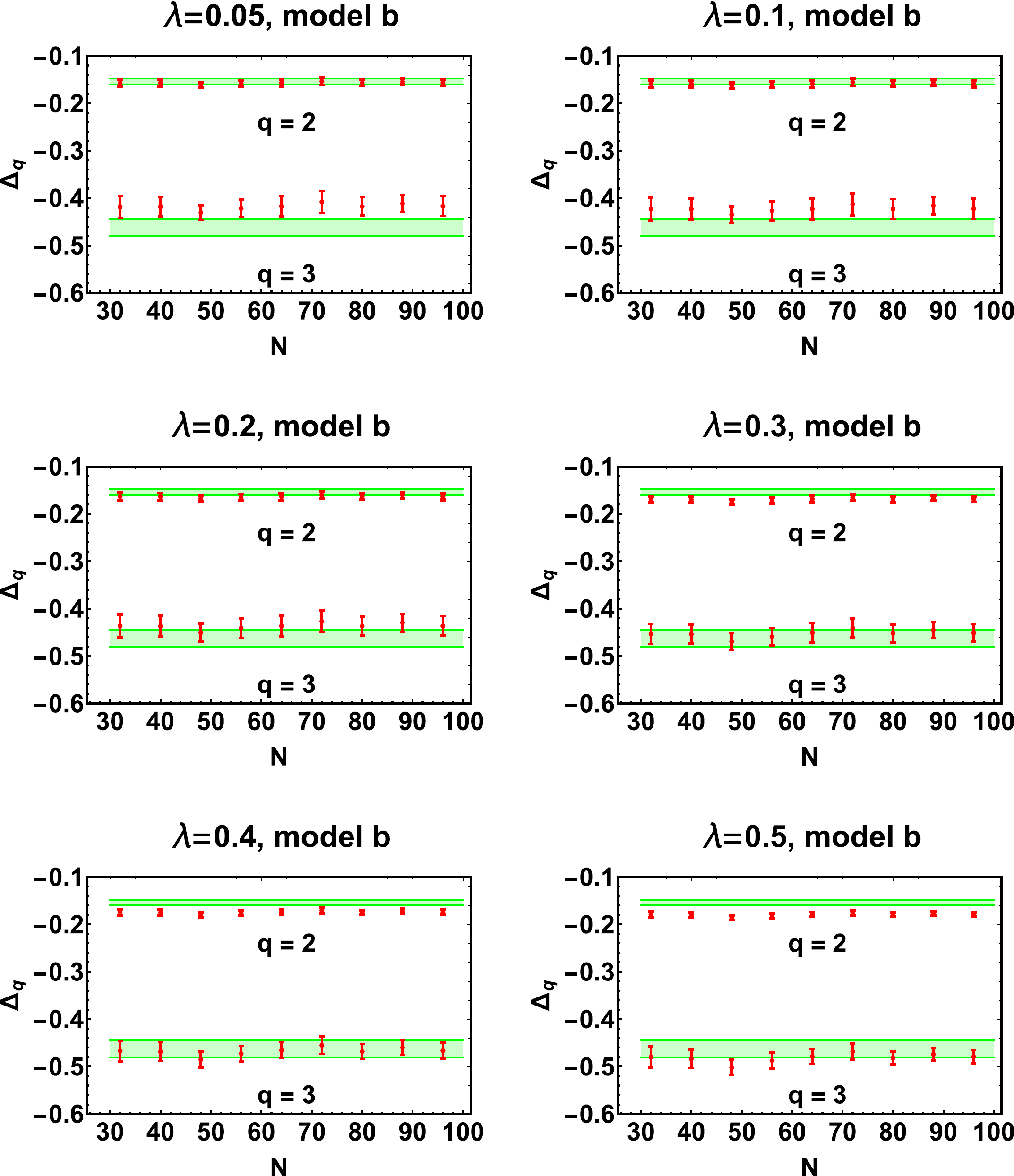}
\caption{Finite-size ($N$) scaling of two particular anomalous multifractal dimensions $\Delta(2)$ and 
$\Delta(3)$ for model (b). In each case the data (red dots) corresponds to the average of an energy bin selected
with the highest percentage of states matching the quadratic ansatz for $\Delta(q)$, shown in Fig.~\ref{Fig--Dirt_Series_Dq(b)}. 
The red error bars indicate the standard deviation of the states in the energy bin.
The green shaded region indicates 
a narrow range of parabolic ans\"atze for $\Delta(q)$,
with $q_c = 5.1 + \delta_c$ ($|\delta_c| \leq 0.1$).}
\label{Fig--N_Series(b)}
\end{figure}

\begin{figure}[t!]
\centering
\includegraphics[width=0.47\textwidth]{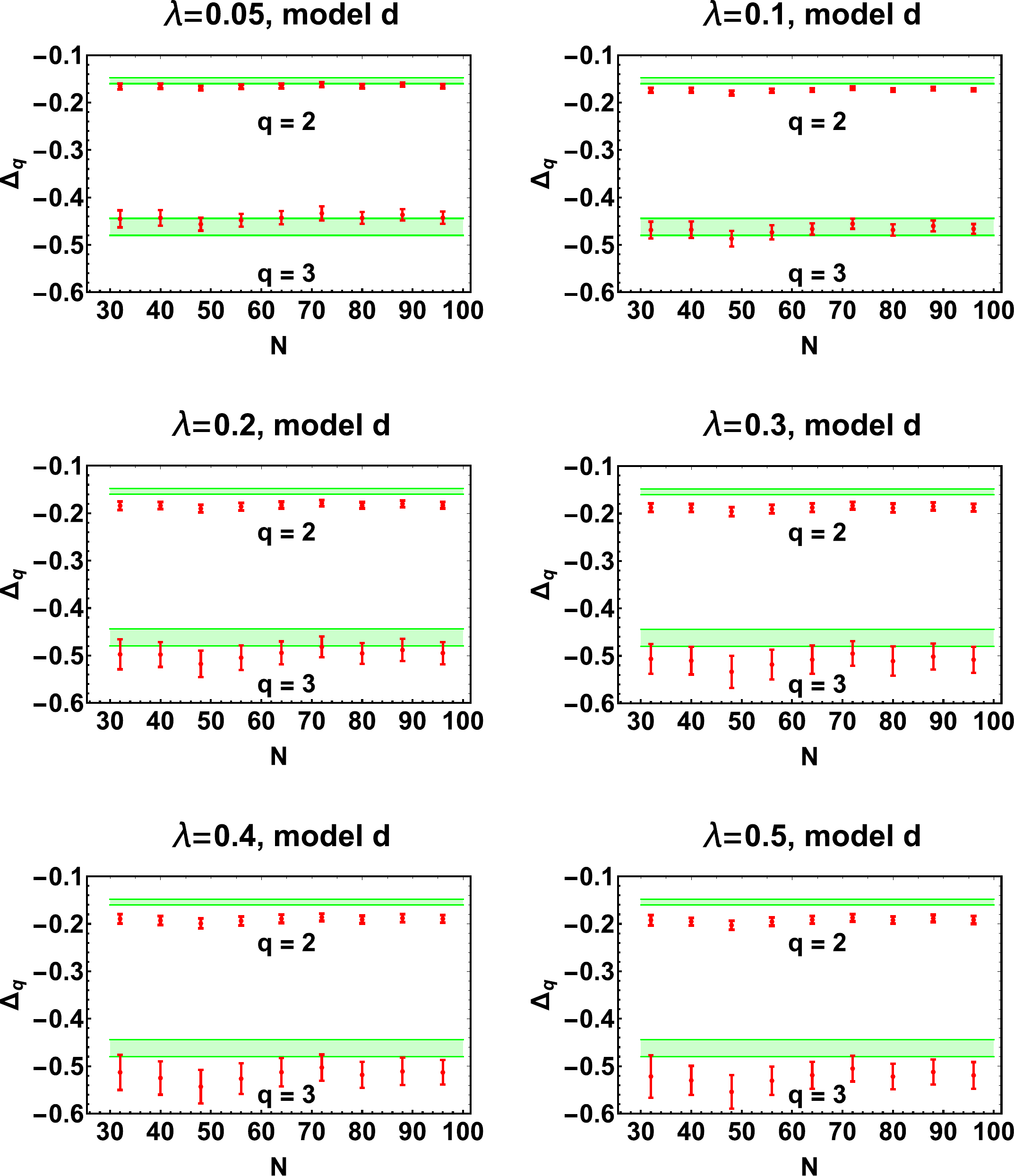}
\caption{Finite-size ($N$) scaling of two particular anomalous multifractal dimensions $\Delta(2)$ and 
$\Delta(3)$ as in Fig.~\ref{Fig--N_Series(b)}, but here computed for model (d). 
The data (red dots) corresponds to the average of an energy bin selected
with the highest percentage of states matching the quadratic ansatz for $\Delta(q)$, shown in Fig.~\ref{Fig--Dirt_Series_Dq(d)}.}
\label{Fig--N_Series(d)}
\end{figure}


\subsection{System-size scaling of the DOS and DOCS \label{Sec--SysSize}}

As emphasized above, weak QGD is strongly irrelevant near the Dirac point. 
For an electrically charged Dirac field $\psi$, kinetic theory predicts 
a divergent dc conductivity for all nonzero temperatures \cite{DavisFoster}. 
The finite-energy critical-state stack found here instead suggests a finite dc conductivity.

On the other hand, conventional symmetry arguments imply that the finite-energy states
of a class DIII model should reside in the symplectic class AII \cite{Evers2008}, see Appendix \ref{Sec:Sym}. 
This class can exhibit weak antilocalization, leading to a ``supermetallic'' phase in 2D \cite{Bardarson07,Nomura07}.
In a finite-size system, it is therefore not surprising
to find multifractal wave functions in 2D, since the scaling is only logarithmic
in the system size. 
However, the curvature of the multifractal spectrum in that case should be related
to the bare disorder strength. 
Moreover, more of the spectrum should become more weakly multifractal
with increasing system size. 

The leading order anomalous multifractal spectrum for a 2D symplectic metal is given by \cite{HoefWegner86}
\begin{align}\label{AII_Delta}
	\Delta(q)
	\equiv
	\tau(q) 
	- 
	2(q-1)
	=
	\frac{1}{8 \pi^2 G(L)}
	q(1-q),
\end{align}
where the dimensionless conductance $G(L)$ scales according to weak antilocalization \cite{LeeRamakrishnan}
\begin{align}\label{AII_G(L)}
	G(L)
	=
	G_0
	+
	\frac{1}{2 \pi^2}
	\log\left(\frac{L}{l_{\mathsf{el}}}\right).
\end{align}
Here $L$ is the system size and $l_{\mathsf{el}}$ denotes the elastic impurity scattering length. 
The bare conductance $G_0 = \nu(\e) D(\e)$, where $\nu(\e)$ is the density of states and $D(\e)$ is the
diffusion constant, at the single-particle energy $\e$. 
Fixing the disorder strength fixes $G_0$ at each energy. Increasing the system size $L$ should then 
reduce multifractality according to Eq.~(\ref{AII_Delta}), as the system flows slowly towards a ``supermetallic''
phase at infinite $L$.

We plot the evolution of the density of states (DOS) and density of critical states (DOCS) as a function
of system size 
for models (b), (d), and (e) in Figs.~\ref{Fig--NSeriesDOCS(b)}--\ref{Fig--NSeriesDOCS(e)}. 
These plots exhibit no discernible systematic variation with system size.
This is not consistent with weak antilocalization behavior expected from the symplectic metal class AII in Eqs.~(\ref{AII_Delta}) and (\ref{AII_G(L)}). 

Finally, in Figs.~\ref{Fig--N_Series(b)} and \ref{Fig--N_Series(d)}, for the robust stacking models (b) and (d) we exhibit the finite-size 
scaling of two particular multifractal dimensions with system size.
Results are shown for the same 6 disorder strengths $\lambda$ employed in 
Figs.~\ref{Fig--DOS(b)}, 
\ref{Fig--DOS(d)}, 
\ref{Fig--Dirt_Series_Dq(b)}, 
and 
\ref{Fig--Dirt_Series_Dq(d)}.
At each system size $N$, the dimensions are computed from an energy bin wherein the DOCS is maximized relative to the DOS. 
These figures indicate a possible very weak dependence of the dimensions on the disorder strength. 
This is very different from the symplectic metal prediction in Eqs.~(\ref{AII_Delta}) and (\ref{AII_G(L)}),
which would instead imply a linear dependence of the dimensions on the disorder strength 
(proportional to $1/G_0$).

\subsection{Discussion and open questions}

We have uncovered a third 
instance of quantum-critical wave function stacking \cite{Ghorashi18}, 
due here to QGD.
In both Ref.~\cite{Ghorashi18} (class CI) and the present paper (class DIII), the stacking occurs 
in disordered Dirac models that can describe the 2D surface states of 3D topological superconductors. 
Another instance (for topological superconductor surface states in class AIII) was touched upon in Ref.~\cite{Chou2014},
and extensively explored in Ref.~\cite{Sbierski}. 
 
We argued that QGD naturally obtains whenever 2D Dirac quasiparticles arise from a spatially inhomogeneous gap.
The critical-state stack found in this paper might provide a simple explanation
for the division between plane-wave-like and spatially inhomogeneous LDOS fluctuations measured
at low and finite energies, respectively, in the high-$T_c$ cuprates. 

We studied the five model variants (a)--(e), defined below Eq.~(\ref{DiracRVH}).
We found the best qualitative agreement with cuprate STM phenomenology \cite{DavisReview} 
for model (b), which includes random pseudospin rotations
and quenched nematic fluctuations of the Dirac cone. Robust stacking that increases with increasing disorder
strength is observed in both model (b) and model (d), the latter which includes only pure nematic disorder. 
By contrast, isotropic flattening or steepening of the Dirac cone appears to suppress critical stacking
for intermediate and strong disorder [models (a) and (c)]. The generic model (e) shows intermediate behavior,
with a robust critical stack at all disorders, but low-energy states that become more rarified (supercritical)
with increasing disorder strength.

Avenues and questions for future work include the following. 
(1) The anomalous multifractal spectrum for spatial LDOS fluctuations computed for the critical stack,
exhibited in 
Figs.~\ref{Fig--Dirt_Series_Dq(b)}--\ref{Fig--Dirt_Series_Dq(e)}, could be directly compared to corresponding spectra
computed from experimental STM data on BSCCO \cite{DavisReview}. 
(2) A crucial question is whether the crossover scale $\Delta_0$ 
[Fig.~\ref{Fig--Delta0vsDirt} and Eq.~(\ref{Delta_0-Scaling})] 
remains finite or vanishes in the infinite system-size $L \rightarrow \infty$ limit. 
A nonzero $\Delta_0(L\rightarrow\infty)$ would indicate a finite-energy transition
between ballistic and quantum-critical stacked behavior.
(3) Another question regards transport versus temperature in the Dirac model with QGD. 
In particular, could the transition from ``ballistic'' low-energy states to critical, finite-energy
ones explain the ubiquitous linear-in-$T$ resistivity observed in the 
strange metal phase above $T_c$ \cite{LeeMott06}? 
(4) What is the role of \emph{dephasing} on transport \cite{AAK,IQHP-Deph}
due to inelastic quasiparticle scattering, which is presumably important 
in the cuprates due to the strong correlations ($U \gg t$), and finally 
(5) Does the multifractal-stacking phenomenon enhance superconductivity in a self-consistent calculation of the gap?

\begin{acknowledgments}

We thank 
Alex Altland,
Mustafa Amin, 
Seamus Davis, 
Ilya Gruzberg, 
Victor Gurarie, 
Christopher Hooley, 
Enrico Rossi, 
and 
Bj\"orn Sbierski for useful discussions.
S.A.A.G.\ 
was supported by the U.S.\ Army Research Office 
Grant No.~W911NF-18-1-0290, 
and acknowledges partial support from 
NSF CAREER 
Grant No.~DMR-1455233
and 
ONR 
Grant No.~ONR-N00014-16-1-3158.
J.F.K.\ acknowledges
the KHYS research travel grant and
Graduate Funding from the German States.
S.M.D.\ and M.S.F.\ acknowledge support 
by the Welch Foundation Grant No.~C-1809 
and 
by NSF CAREER Grant No.~DMR-1552327.
M.S.F.\ also acknowledges support 
by the U.S. Army Research Office Grant No.~W911NF-17-1-0259.
M.S.F.\ thanks the Aspen Center for Physics, which is
supported by the NSF Grant No.~PHY-1607611, for its
hospitality while part of this work was performed.

\end{acknowledgments}

\appendix


\begin{widetext}

\section{Random velocity modulation from Dirac fermions in curved spacetime \label{Sec:Gravity}}

The generally covariant action for (2+1)-D Dirac fermions propagating through curved spacetime
is given by Eq.~(\ref{DiracCurv}). In the context of disordered Dirac superconductors with QGD,
however, there is a naturally preferred coordinate system, where $\vex{r} = \{x_1,x_2\}$ measure
physical Euclidean distances across the sample. In this ``physical'' coordinate system,
the Berry phase term $\bar{\psi}\gh^0 i \parr_t \psi$ arises from Trotterization
in the path integral, and should therefore have unit coefficient. We call this
the ``temporal flatness condition.'' Then, we can take the dreibein $E_A^\mu$ in Eq.~(\ref{DiracCurv}) 
to be off-diagonal only in the spatial-spatial sector. Let $i,j \in \{1,2\}$. 
We define 
\begin{align}\label{vijDreibeinDef}
	v_{i j}(\vex{r}) \equiv \frac{E_i^j(\vex{r})}{E_0^0(\vex{r})}.
\end{align}
Temporal flatness requires 
\begin{align}\label{TFC}
	\sqrt{|g|} \, E_0^0 = 1.
\end{align}
Shifting $\bar{\psi} \rightarrow \bar{\psi} \gh^0$ and enforcing Eqs.~(\ref{vijDreibeinDef}) and (\ref{TFC}),
Eq.~(\ref{DiracCurv}) reduces to Eq.~(\ref{DiracRV}).
The spin connection is eliminated via integration-by-parts in the spatially modulated kinetic terms;
this leads to the manifestly Hermitian form of the Hamiltonian in Eqs.~(\ref{DiracRVH}).

The inverse metric can be expressed as 
\begin{align}
	g^{\mu \nu}
	=	
	E^{\mu}_A \, E^\nu_B \, \eta^{A B}
	\rightarrow 
	\frac{1}{\sqrt{D}}
	\begin{bmatrix}
	-1 			& 0 				& 0 				\\
	0 			& \voo^2 + \vto^2	 	& \voo \, \vot + \vto \, \vtt		\\
	0 			& \voo \, \vot + \vto \, \vtt		& \vot^2 + \vtt^2
	\end{bmatrix},
\end{align}
where $\eta^{A B} \rightarrow \diag(-1,1,1)$ is the Minkowski metric, 
and
\begin{align}
	D \equiv \frac{1}{\left(E_0^0\right)^4}.
\end{align}
Imposing Eq.~(\ref{TFC}), the metric is  
\begin{align}
	g_{\mu \nu}
	\rightarrow 
	\frac{1}{\sqrt{D}}
	\begin{bmatrix}
	-D 			& 0 				& 0 					\\
	0 			& \vtt^2 + \vot^2	 	& -\voo \, \vot - \vto \, \vtt		\\
	0 			& -\voo \, \vot - \vto \, \vtt	& \voo^2 + \vto^2
	\end{bmatrix},
	\qquad
	D 
	=
	\left(\voo \, \vtt - \vot \, \vto\right)^2. 
\end{align}
This can be specialized for the five model variants (a)--(e), defined below Eq.~(\ref{DiracRVH}).
For example, model (a) has $\vot = \vto = 0$, leading to 
\begin{align}
	g_{\mu \nu}^\puprm{a}
	\rightarrow 
	\frac{1}{\voo \, \vtt}
	\begin{bmatrix}
	-\voo^2 \, \vtt^2 	& 0 				& 0 					\\
	0 			& \vtt^2	 		& 0					\\
	0 			& 0				& \voo^2
	\end{bmatrix},
\end{align}
while model (b) has $\voo = \vtt = 1$, and 
\begin{align}
	g_{\mu \nu}^\puprm{b}
	\rightarrow 
	\frac{1}{1 - \vot \, \vto}
	\begin{bmatrix}
	-(1 - \vot \, \vto)^2	& 0 				& 0 					\\
	0 			& 1 + \vot^2	 		& -\vot - \vto				\\
	0 			& -\vot - \vto			& 1 + \vto^2
	\end{bmatrix}.
\end{align}

In the generic case, the scalar curvature can be written has 
\begin{align}\label{RScalar}
	R 
	=
	\frac{N}{2\left(\voo \, \vtt - \vot \, \vto\right)^3},
\end{align}
where $N$ is a homogeneous quadratic in spatial derivatives of $\{v_{i j}\}$. 
As an example, for model (a) with $\voo = \vtt \equiv v(\vex{r})$, 
\begin{align}
	R^\puprm{a}
	=
	- 
	2 
	v^{-1}
	\nabla^2
	v.
\end{align}

\end{widetext}


\section{``Gravitational'' coupling of electric potentials to the surface Majorana fluid of a class DIII topological superconductor \label{Sec:3HeB}}

In this section we derive the form of the velocity modulation in Eq.~(\ref{vDIII}),
corresponding to the effect of an electric potential $A^0(\vex{r})$ on the 2D Majorana
fluid expected to form at the surface of a class DIII topological superconductor.

\subsection{Bulk and surface states for solid-state $^3$He-$B$}

As a simple model for a class DIII bulk topological superconductor with winding number $|\nu| = 1$, 
we consider a solid-state analog of $^3$He-$B$ \cite{Volovik,TSCRev2}.
The bulk Bogoliubov-de Gennes Hamiltonian features
isotropic $\sigb\cdot\vex{k}$ pairing, where $\sigb = \sigh^a \hat{x_a}$ is the vector of Pauli matrices
acting on the physical spin-1/2 components $\sigma \in \{\uparrow,\downarrow\}$, and $a$ is summed over $\{1,2,3\}$.  
The mean-field Hamiltonian is 
\begin{align}\label{Ham}
\begin{aligned}[b]
	H 
	=&\,
	\frac{1}{2}
	\intl{\vex{k}}
	\chi^\dagger(\vex{k})
	\,
	\hs(\vex{k}) 
	\,
	\chi(\vex{k}),
\\
	\hs(\vex{k})
	=&\,
	\left(
		\frac{k^2}{2m} 
		- 
		\mu
	\right)
	\tauh^3 
	+
	\Delta
	\left(\sigb\cdot\vex{k}\right) 
	\tauh^2,		
\end{aligned}
\end{align}
where $\mu > 0$ is the chemical potential and $\Delta$ is the $p$-wave pairing amplitude. 
Here we have introduced the Balian-Werthammer (``Majorana'') spinor
\begin{align}\label{MajDef}
	\chi(\vex{k}) 
	\equiv&\,
	\begin{bmatrix}
	c(\vex{k})	\\
	\sigh^2 \left[c^\dagger(-\vex{k})\right]^\T
	\end{bmatrix},
\;\;
	\chi^\dagger(\vex{k}) 
	=
	i 
	\,
	\chi^\T(-\vex{k}) 
	\Mp.
\end{align}
The Pauli matrices $\{\tauh^{1,2,3}\}$ act on particle-hole space. 
Particle-hole $P$, time-reversal $T$, and chiral $S$ ($\equiv T \times P$) 
symmetries are defined via
\bsub
\label{PTS}
\begin{align}
\begin{aligned}
P:&\,&
	- \Mp^{-1} \, \hs^\T(-\vex{k}) \, \Mp =&\, \hs(\vex{k}), 
\\
T:&\,&
	\Mt^{-1} \, \hs^*(-\vex{k}) \, \Mt =&\, \hs(\vex{k}), 
\\
S:&\,&
	- \Ms \, \hs(\vex{k}) \, \Ms =&\, \hs(\vex{k}), 
\end{aligned}
\end{align}
where
\begin{align}
\begin{aligned}
	\Mp=&\, \sigh^2 \tauh^2 = \Mp^\T, 
	\;\;
	&\,
	(P^2 = +1),
\\
	\Mt=&\, i \sigh^2 \tauh^3 = - \Mt^\T, 
	\;\;
	&\,
	(T^2 = -1),
\\
	\Ms=&\, \tauh^1,
	\;\;
	&\,
\end{aligned}
\end{align}
\esub
consistent with class DIII \cite{SRFL08,Foster14}. 

As in \cite{Ghorashi17}, we implement hardwall boundary conditions 
at $z = 0$ in order to get surface states. 
Eq.~(\ref{Ham}) separates into a $\vex{k} = 0$ piece, and a nonzero $\vex{k}$ piece,
where $\vex{k} \equiv \{k_x,k_y\}$ now accounts only for conserved transverse momenta. 
\begin{align}\label{Ham2}
\begin{aligned}
	\hs
	=&\,
	\hs_0
	+
	\hs_1,
\\
	\hs_0
	=&\,
	\left(
		-
		\frac{1}{2m}
		\partial_z^2
		- 
		\mu
	\right)
	\tauh^3 
	+
	\Delta 
	(-i \partial_z)
	\sigh^3
	\tauh^2,
\\ 
	\hs_1
	=&\,
	\left[\frac{k^2}{2m} - e A^0(\vex{r})\right]
	\,
	\tauh^3 
	+
	\Delta
	\left(\sigb\cdot\vex{k}\right)
	\tauh^2.
\end{aligned}
\end{align}
We have included a scalar electric potential $A^0(\vex{r})$ as a perturbation in $\hs_1$. 

The Hamiltonian $\hs_0$ has a pair of zero-energy Majorana bound states,
\begin{align}
\label{ZM}
	\ket{\psi_{0,m_s}} 
	=
	\ket{\tau^1 = m_s}
	\otimes
	\ket{m_s}
	\otimes
	\ket{\varphi},
\end{align}
where $\ket{\tau^1 = m_s}$ is the particle-hole ($\tau$) space spinor,
which is ``locked'' to the $\sigh^z$-spin projection $m_s$ (in the plus and minus $\tauh^1$-direction
for $m_s = \uparrow$ and $\downarrow$, respectively). 
The spatial profile of the bound state is 
\begin{align}
\label{fEval}
	\braless{z}\!
	\ket{\varphi}
	=
	\frac{1}{\sqrt{\nm{0}}}
	\,
	e^{ - m \Delta z}
	\sin\left[z \sqrt{2 m \mu - m^2 \Delta^2}\right],
\end{align}
where $\nm{0}$ is a normalization factor.

\subsection{$k \cdot p$ perturbation theory}

An effective Hamiltonian for the surface theory at $\vex{k} \neq 0$ obtains by taking matrix elements of 
the following operator between the $m_s = \pm 1$ zero modes in Eq.~(\ref{ZM}),
\begin{align}\label{Heff}
	\hsS
	=&\,
	\hs_1
	-	
	\hs_1
	\,
	\hat{P}_1
	\,
	\hs_0^{-1}
	\,
	\hat{P}_1 
	\,
	\hs_1
	+
	\ldots
\nonumber\\
	=&\,
	\hs_1
\nonumber\\
	&\,
	-	
	\sum_{m_s = \pm 1}
	\int_0^\infty 
	\frac{d q}{2 \pi E_q}
	\,
	\hs_1
	\left[
	\begin{aligned}
	&\,
		\ket{\psi_{q,m_s}}\bra{\psi_{q,m_s}}
	\\&\,
	-
		\tauh^1 \ket{\psi_{q,m_s}}\bra{\psi_{q,m_s}} \tauh^1
	\end{aligned}
	\right]
	\hs_1
\nonumber\\
	&\,
	+
	\ldots,
\end{align}
where we have expanded the second term via the $\vex{k} = 0$ resolution of the identity.
The operator $\hat{P}_1$ on the first line is the projection out of the 
degenerate eigenspace of the zero modes, $\hat{P}_1 = \hat{1} - \hat{P}_0$, where 
\[
	\hat{P}_0
	=
	\sum_{m_s = \pm 1}
	\ket{\psi_{0,m_s}}\bra{\psi_{0,m_s}}.
\]
The state $\ket{\psi_{q,m_s}}$ is a positive-energy (gapped) bulk eigenstate
of $\hs_0$, parameterized by the standing wave momentum $q$, 
while $\tauh^1 \ket{\psi_{q,m_s}}$ is its negative-energy chiral conjugate [Eq.~(\ref{PTS})].  
$E_q = \sqrt{(q^2/2m - \mu)^2 + q^2 \Delta^2}$ denotes the positive eigenenergy. 

The first term in Eq.~(\ref{Heff}) gives the relativistic dispersion for the Majorana surface fluid,
\begin{align}\label{hsSe1}
	\hsS^{\pup{1}}
	=
	\Delta 
	\,
	\sigb
	\wedge
	\vex{k},
\end{align}
where $\vex{A} \wedge \vex{B} = A_x B_y - A_y B_x$. 
This is consistent with the surface projection of the symmetry conditions in Eq.~(\ref{PTS}), 
\bsub
\label{PTS-Surf}
\begin{align}
\begin{aligned}
P:&\,&
	- (\Mps)^{-1} \, \hsS^\T(-\vex{k}) \, \Mps =&\, \hsS(\vex{k}), 
\\
T:&\,&
	(\Mts)^{-1} \, \hsS^*(-\vex{k}) \, \Mts =&\, \hsS(\vex{k}), 
\\
S:&\,&
	- \Mss \, \hsS(\vex{k}) \, \Mss =&\, \hsS(\vex{k}), 
\end{aligned}
\end{align}
where 
\begin{align}
\begin{aligned}
	\Mps=&\, \sigh^1 = (\Mps)^\T 
	\;\;
	&\,
	(P^2 = +1),
\\
	\Mts=&\, i \sigh^2 = - (\Mts)^\T 
	\;\;
	&\,
	(T^2 = -1),
\\
	\Mss=&\, \sigh^3.
	\;\;
	&\,
\end{aligned}
\end{align}
\esub

Coupling to the vector potential $A^0$ obtains from the second term in Eq.~(\ref{Heff}).
Working to linear order in the potential, 
the relevant $+-$ matrix elements take the form
\begin{widetext}
\begin{align}
	-
	&
	\sum_{m_s = \pm 1}
	\int_0^\infty 
	\frac{d q}{2 \pi E_q}
	\,
	\bra{\tau^1 = +1}
	\bra{+1}
	\bra{\varphi}
	\left[
	-
	e
	A^0
	\tauh^3 
	\right]	
	\Big[
		\ket{\psi_{q,m_s}}\bra{\psi_{q,m_s}}
		-
		\tauh^1 \ket{\psi_{q,m_s}}\bra{\psi_{q,m_s}} \tauh^1
	\Big]
\left[
	\Delta
	\,
	\sigb\cdot\vex{k}
	\,
	\tauh^2
	\right]	
	\ket{\tau^1 = -1}
	\ket{-1}
	\ket{\varphi}
\nonumber\\
&\,
	=
	e
	A^0
	\,
	\Delta 
	i 
	\kc
	\int_0^\infty 
	\frac{d q}{\pi E_q}
	\bra{\varphi}
	\braless{\tau^1 = -1}\!
	\ket{\psi_{q,+1}}
	\braless{\psi_{q,+1}}\!
	\ket{\tau^1 = +1}
	\ket{\varphi},
\end{align}
and 
\begin{align}
	-
	&
	\sum_{m_s = \pm 1}
	\int_0^\infty 
	\frac{d q}{2 \pi E_q}
	\,
	\bra{\tau^1 = +1}
	\bra{+1}
	\bra{\varphi}
	\left[
	\Delta
	\,
	\sigb\cdot\vex{k}
	\,
	\tauh^2
	\right]	
	\Big[
		\ket{\psi_{q,m_s}}\bra{\psi_{q,m_s}}
		-
		\tauh^1 \ket{\psi_{q,m_s}}\bra{\psi_{q,m_s}} \tauh^1
	\Big]
	\left[
	-
	e
	A^0
	\tauh^3 
	\right]	
	\ket{\tau^1 = -1}
	\ket{-1}
	\ket{\varphi}
\nonumber\\
&\,
	=
	\Delta 
	i
	\kc
	\,
	e
	A^0
	\int_0^\infty 
	\frac{d q}{\pi E_q}
	\,
	\bra{\varphi}
	\braless{\tau^1 = -1}\!
	\ket{\psi_{q,-1}}
	\braless{\psi_{q,-1}}\!\ket{\tau^1 = +1}\ket{\varphi},
\end{align}
\end{widetext}
where $\kc \equiv k_x - i k_y$. 
Evaluating these leads to perturbation of the form
\begin{align}\label{hsSe2}
	\hsS^{\pup{2}}
	=
	\frac{\vartheta \, \Delta}{2}
	\left[
		\frac{e A^0(\vex{r})}{E_{\mathsf{bulk}}}
		\,
		\sigh
		\wedge
		\vex{k}
		+
		\sigh
		\wedge
		\vex{k}
		\,
		\frac{e A^0(\vex{r})}{E_{\mathsf{bulk}}}
	\right],
\end{align}
where $E_{\mathsf{bulk}} \simeq k_F \Delta$ is the bulk excitation gap, and $\vartheta$ is a pure order-one number. 
Since the bare Majorana fluid velocity is $\Delta$, we recover Eq.~(\ref{vDIII}).


\section{Symmetry class for finite-energy states; Connection to the class D thermal quantum Hall plateau transition (?) \label{Sec:Sym}}

The 2D velocity-randomized Dirac Hamiltonian given by the sum $\hsS^{\pup{1}} + \hsS^{\pup{2}}$ 
in Eqs.~(\ref{hsSe1}) and (\ref{hsSe2}) resides in class DIII, due to the $T^2 = -1$ and $P^2 = +1$ 
time-reversal and particle-hole symmetries encoded in Eq.~(\ref{PTS-Surf}). 
These symmetries hold irrespective of whether the fermion field $\psi$ in Eq.~(\ref{DiracRV}) 
is a complex-valued Dirac or real-valued
Majorana spinor. 

Typically, we can associate an effective field theory, the nonlinear sigma model, to describe the 
wave functions of any single-particle Hamiltonian at some particular fixed energy. 
The nonlinear sigma model employs local operators to encode the probability statistics of the extended, critical, or localized states. 
Using fermionic replicas to perform disorder-averaging, class DIII is associated to a sigma model
with the target manifold O$(2 n)$, where $n$ is proportional to the number of replicas \cite{Evers2008}. 
In two spatial dimensions, this can be seen via the nonabelian bosonization of the clean, zero-energy Majorana 
fermion field theory \cite{Foster14}. 

For the DIII theory, nonzero energy is a relevant perturbation that couples to the principal chiral field \cite{Foster14,Ghorashi18}
in the nonlinear sigma model. In this case, the O($2n$)$\times$O($2n$) symmetry of the zero-energy theory is broken down to the
diagonal subgroup O($2n$). The energy perturbation will induce an RG flow to a new fixed point, which should be associated
to a different sigma model with target manifold O($2n$)/$H$. There are only two possibilities: 
\bsub\label{DIII-AII-D}
\begin{enumerate}
\item{Class AII, the ``symplectic'' class associated to disordered metals with time-reversal symmetry and strong spin-orbit coupling. 
This class typically exhibits weak antilocalization in two dimensions. 
Class AII also describes the 2D surface states of 3D topological insulators. The target manifold is \cite{Evers2008}
\begin{align}
	G/H = \frac{\text{O}(2n)}{\text{O}(n)\,\times\,\text{O}(n)}.
\end{align}
}
\item{Class D, typically associated to superconductors with broken time-reversal symmetry and strong spin-orbit coupling. 
Class D should be realized in dirty $p+ip$ superconductors \cite{D-1,D-2,D-5,D-6,D-7}. The target manifold is 
\begin{align}
	G/H = \frac{\text{O}(2n)}{\text{U}(n)}.
\end{align}
}
\end{enumerate}
\esub

We argued in Sec.~\ref{Sec:Results} that class AII appears incompatible with the numerical
results obtained here for finite-energy states of the velocity-modulated Dirac Hamiltonian in Eq.~(\ref{DiracRVH}). 
We therefore expect that our finding of critical states throughout the energy spectrum with universal multifractal
spectra should instead be associated to class D. 

There are three classes of time-reversal invariant topological superconductors (TSCs) in 3D,
differing by the amount of spin rotational symmetry: 
CI [SU(2)], AIII [U(1)], and DIII (no spin symmetry, strong spin-orbit coupling) \cite{ClassRevNote}. 
There are also three classes of time-reversal broken quantum Hall topological insulators or TSCs in 2D: 
class C [spin quantum Hall (SQH) effect], 
class A [ordinary integer quantum Hall (IQH) effect], 
and 
class D [thermal quantum Hall (TQH) effect] \cite{Evers2008}. 
All of these classes are characterized by integer (AIII, DIII, A, D) or twice-integer (CI, C) bulk topological winding numbers \cite{ClassRevNote}.   

In our previous work \cite{Ghorashi18}, we performed a similar population analysis as presented in this paper 
for the finite-energy 2D surface states of a 3D class CI topological superconductor.
Based on the numerical results obtained there, 
we concluded that the finite-energy states of the 2D class CI model take the form of a ``stack''
of critical wave functions with universal multifractal spectra. The spectra are energy independent,
and consistent with the plateau transition of the spin quantum Hall effect in class C
\cite{SQHP3,EversSQHE03,MirlinSQHE03}. 
A connection between classes CI and C based on symmetry considerations similar to Eq.~(\ref{DIII-AII-D})
was also presented in Ref.~\cite{Ghorashi18}. 

Finite-energy states in class AIII must reside in class A \cite{Ludwig1994}.
For the 2D surface states of a 3D class AIII topological superconductor,
whether these are critically delocalized or Anderson localized
depends upon the presence or absence of a topological theta term (at $\theta = \pi$) in the effective nonlinear sigma model \cite{Pruisken}. 
In Ref.~\cite{Sbierski}, strong numerical evidence was presented that 2D finite-energy surface states of 3D class AIII TSCs 
form a ``stack'' of critical wave functions [see also Ref.~\cite{Chou2014}]. 
The  spectra are energy independent, 
and consistent with the plateau transition of the ordinary class A integer quantum Hall effect \cite{Huckestein1995}. 
For classes CI and AIII, the only alternative to the plateau-transition-stacking scenario is Anderson localization,
but this is not observed in the numerical studies \cite{Ghorashi18,Chou2014,Sbierski}.

We therefore expect that the ``stack'' of critical states found in the present paper
can be associated to the thermal quantum Hall plateau transition in class D. 
In comparison to classes C and A, relatively little is known about the thermal quantum Hall plateau transition.
The global phase diagram for a 2D system in class D is complicated by the advent of a thermal metal phase \cite{D-1,D-2,D-5,D-6,D-7}, in addition
to Anderson localized thermal Hall plateaux \cite{Evers2008}.  
Further studies of the possible \emph{multicritical point} in the phase diagram of class D \cite{D-1,D-5,D-6} 
could shed light on the nature of the finite-energy states found here.


\end{document}